\documentclass[onecolumn,authoryear]{els-mrw} 

\usepackage{amsmath,amssymb,amsfonts,amsthm,makeidx,graphicx}
\usepackage{txfonts}
\usepackage{helvet}
\usepackage{graphicx}
\usepackage{blindtext}

\makeatletter
\renewcommand\tableofcontents{%
  \@starttoc{toc}%
}
\makeatother

\begin{document}

\chapter{Giant branch systems: surveys and populations}\label{chap1}

\author[1,2]{Samuel Kai Grunblatt}%
%\author[2]{Second Author}%

%\author[1,2]{Third Author}%

\address[1]{\orgname{University of Alabama}, \orgdiv{Department of Physics and Astronomy}, \orgaddress{Tuscaloosa, AL 35487}}
\address[2]{\orgname{Johns Hopkins University}, \orgdiv{William H. Miller III Department of Physics and Astronomy}, \orgaddress{3400 N. Charles St., Baltimore, MD 21218}}
%\address[2]{\orgname{Name of Institute}, \orgdiv{Division or Department}, \orgaddress{Address of Institute}}

\articletag{1st edition.}

\maketitle

\begin{keywords}[Keywords]

giant stars, planetary system evolution, star-planet interactions, galactic archaeology, exoplanet atmospheric evolution

\end{keywords}
%
%\begin{frontmatter}
%\end{frontmatter}
%\\
%\tableofcontents
\section*{Contents}
\contentsline {section}{\numberline {1}Introduction}{2}{}%
\contentsline {section}{\numberline {2}Radial velocity observations: the first discoveries}{2}{}%
\contentsline {subsection}{\numberline {2.1}``Retired" A star surveys: key to determining planet occurrence as a function of stellar mass}{4}{}%
\contentsline {section}{\numberline {3}Transit surveys}{5}{}%
\contentsline {subsection}{\numberline {3.1}Breakthroughs by accident: the Kepler era}{6}{}%
\contentsline {subsection}{\numberline {3.2}Targeted surveys: \emph {K2} and \emph {TESS}}{7}{}%
\contentsline {subsection}{\numberline {3.3}Future surveys}{7}{}%
\contentsline {paragraph}{PLATO}{7}{}%
\contentsline {paragraph}{Roman}{7}{}%
\contentsline {section}{\numberline {4}Evolved Planet Populations}{7}{}%
\contentsline {subsection}{\numberline {4.1}Hot Jupiters}{8}{}%
\contentsline {paragraph}{Overcoming early biases}{9}{}%
\contentsline {paragraph}{Orbital processes: laboratories}{10}{}%
\contentsline {subparagraph}{Circularization and Inspiral: a population perspective}{10}{}%
\contentsline {subparagraph}{Spin-orbit obliquity}{10}{}%
\contentsline {paragraph}{Atmospheric processes: laboratories}{11}{}%
\contentsline {subparagraph}{Re-inflated planets}{12}{}%
\contentsline {subparagraph}{Atmospheric mass loss (in evolved systems)}{13}{}%
\contentsline {subsection}{\numberline {4.2}Longer period planets}{14}{}%
\contentsline {subsection}{\numberline {4.3}Multiplanet systems}{15}{}%
\contentsline {subsection}{\numberline {4.4}Evidence for planet engulfment}{15}{}%
\contentsline {subsection}{\numberline {4.5}Future of evolved planet population studies: planets across the Galaxy}{15}{}%
\contentsline {section}{\numberline {5}Conclusion}{17}{}%
\contentsline {section}{\numberline {6}Further information}{17}{}%

\begin{glossary}[Glossary \& Nomenclature]
\term{Circularization} The slow transition from an eccentric to circular planetary orbit.

\term{Evolved star} Star which has evolved off of the main sequence, typically in the last 10\% of its lifetime. Evolved stars can span a wide range of luminosities but a rather narrow range in temperature. Evolved stars can also sometimes refer to white dwarfs, the final stage of a Sunlike star's lifetime.

\term{Exoplanet} Planet orbiting a star other than our Sun.

\term{Hot Jupiter} Planet with a mass and size similar to Jupiter orbiting its host star with an average orbital period of 10 days or less.

\term{Inspiral} The shrinking of a planet's orbit over astrophysical timescales.

\term{Main sequence} A continuous sequence in stellar temperature and luminosity which represents all stars at all masses at some time. Stars fall very close to this sequence for the vast majority of their lifetimes.

\term{Planetary re-inflation} During the post-main sequence evolution of a planetary host star, gaseous planets that have been shrinking and cooling over time begin to heat up as their host star grows, causing the planetary atmospheres to grow and re-inflating the size of the planet to its natal state.

\term{Post-main sequence star} Synonym for evolved star.

\term{Red giant branch star} Star which has evolved off of the main sequence and through the subgiant branch and is beginning to increase in brightness and radius as the star enters into the last 10\% of its life. 

\term{Spin-orbit obliquity} The angle between the spin axis of a star and the orbital plane of a planet. 

\term{Subgiant} Star which has just evolved off of the main sequence and is cooling significantly while its radius and luminosity is relatively constant. Once the star begins to grow in radius while remaining relatively unchanged in effective temperature, it has left the subgiant phase and begun red giant branch evolution.

%\term{Hot start} 

%\term{Intermediate-mass star} Star with a mass between 1 and 2 solar masses.

%\term{Asteroseismology} The study of the oscillations of stars, often to determine stellar fundamental properties such as mass and radius.

\end{glossary}
%
%\begin{glossary}[Nomenclature]
%\begin{tabular}{@{}lp{34pc}@{}}
%%AF &Assessment Factor\\
%%ECHA &European Chemical Agency\\
%%EPM &Equilibrium Partitioning Method Equilibrium Partitioning Method Equilibrium Partitioning Method Equilibrium\hfill\break Partitioning Method\\
%%ERA &Ecological Risk Assessment\\
%%HC &Hazardous Concentration\\
%
%
%PR & Planetary Re-inflation\\
%
%
%\end{tabular}
%\end{glossary}

\begin{abstract}[Abstract]
Despite the recent discoveries of planets orbiting stars at all evolutionary stages, the evolution of planetary systems remains poorly understood. Studying planetary systems around red giant branch stars can reveal how main sequence planetary systems can change and evolve into white dwarf systems over time. Decades of radial velocity and transit surveys have yielded the detection of hundreds of planets and planet candidates orbiting evolved stars. These planetary systems have provided important insights into understanding how planetary atmospheres and orbits can be disrupted by stellar evolution, potentially being restructured at late stages, and how planets can be eventually engulfed by their stars, possibly reborn as white dwarf planetary systems. Evolved star targets will reveal planet occurrence at the largest distances and most varied environments across the Galaxy.

%Recent studies of the demographics of these systems have revealed that the orbital configurations of these systems differ from those of main sequence systems, where evolved planetary systems seem to display unique orbital and chemical abundance trends not seen in main sequence systems. A holistic analysis of this population will clarify the connections between star and planet evolution, and can also provide new insights into planet demographics that are not easily understood in main sequence systems, such as the occurrence of planets around intermediate-mass (1-2 M$_\odot$) stars. Continued followup of these systems will eventually bridge the gap between main sequence and white dwarf planetary systems, providing a complete picture of planet demographics over all evolutionary stages of a planetary system.
\end{abstract}

\begin{BoxTypeA}[chap1:box1]{Key points}
\begin{itemize} 
\item Planets orbiting evolved stars are in the last 10\% of their lives.
\item Due to the rapid structural changes of their host star, short-period planets around evolved stars were originally expected to be rare, while longer-period planets were predicted to survive these later stages of stellar evolution.
\item Despite this, surveys have found a similar number of hot Jupiters around main sequence and early red giant stars. There is also evidence for a comparable number of long-period giant planets around main sequence and evolved stars, in line with predictions.
\item Stars spanning a wide range of masses which appear quite different on the main sequence, appear much more similar during the red giant stage of stellar evolution.
\item Hot Jupiters orbiting evolved stars may show evidence for multiple paths to formation.
\item Planets orbiting evolved stars appear to follow a log-linear period-eccentricity relation, suggestive of a higher frequency of planet-planet scattering and high eccentricity migration events in these systems.
\item Hot Jupiters orbiting evolved stars show tentative evidence for rapid alignment with their host star after post-main sequence evolution begins.
\item Planets orbiting evolved stars often appear to be larger than their main sequence counterparts, suggesting that their atmospheres respond directly to the increase in heating from their host stars through a process known as 're-inflation.'
\item Due to the brightness of their host stars, evolved systems can be detected and compared over a wider range of distances from the Earth than their main sequence counterparts. This makes evolved systems particularly important for understanding planet demographics at the largest distances across the Galaxy. 

%\item 
%\begin{itemize}%
%\item conservation of motion,
%\item conservation of water, and
%\begin{itemize}%
%\item conservation of mass,
%\item conservation of heat,
%\end{itemize}
%\item conservation of other gaseous and aerosol materials.
%\end{itemize}
%\item The heating or cooling of the lowest levels of the atmosphere by
%the bottom surface is of comparatively small magnitude.
\end{itemize}%%
\end{BoxTypeA}

\section{Introduction}\label{chap1:sec1}

\hspace*{10mm} Planetary systems spend the majority of their lifetimes on their main sequence. However, near the end of the life of a Sunlike star, it will cool while growing in size, as its core becomes denser, burning hydrogen in a shell on an ever-growing mass of helium, while its outer envelope becomes less dense. Eventually, this outer envelope will reach the size of Earth's orbit, and then will continue growing until the core underneath it becomes hot enough to ignite helium, resulting in a sudden shrinking and heating of outer envelope of the star. The star then repeats the process of growing larger as its core heats up and shrinks in size until additional fusion processes are triggered, expelling the outer envelope of the stars in shells until all that remains is a white dwarf, a remnant of a giant star's core, and leftover material from the stellar envelope that has fallen back toward the star. Throughout this process, planetary systems must respond to the violent death throes of their host star.

\hspace*{10mm} At first, evolved stars were expected to be planet deserts, devoid of planetary systems due to their rapid evolutionary changes. However, discoveries over time revealed planets existing more and more precariously close to the edge of stability, gradually changing the view of these systems from planet deserts to an area of active planetary system investigation. 

\hspace*{10mm} In the Hertzsprung-Russell diagram shown in Figure 1, the three largest distinct populations of stars which have been identified by the Gaia spacecraft have been labeled: the Main Sequence stars, Red Giant stars, and White Dwarf stars. Planetary material has now been directly detected around stars of all three of these types. It is unclear whether the planetary material seen around white dwarfs has the same origin as the planets observed around main sequence stars, but a better understanding of how planetary systems change during the transitional, red giant phase can help us to determine how to trace planetary material in white dwarf systems back to its main sequence history.

\hspace*{10mm} In addition, planets around evolved stars reveal how planetary systems are reshaped, and are eventually reborn alongside their host stars. The rapid and easily observable changes in the subgiant and giant stages of stellar evolution induce large changes in the planet populations, making evolved systems ideal laboratories for understanding star-planet interaction and orbital dynamics. The difference in dominant stellar evolutionary processes on the main sequence and during post-main sequence evolution results in distinct patterns appearing in the population of planets detected around stars of different evolutionary states, revealing how the specific aspects of a star at a given evolutionary phase affect existing planetary properties.

%The relative stability of stars on the main sequence for gigayears allows for the isolation of the effects that occur during post-main sequence evolution, such as an increase in irradiation and tidal forcing from the star, from processes more common at other evolutionary states, such as magnetic activity.

\hspace*{10mm} Here we outline the history of planet discovery around evolved stars, and its eventual growth into the study of this unique planet population in context of all known planets. We begin with the initial individual planet discoveries, and then discuss broader surveys and our current understanding of the subpopulations of evolved planets that have been found so far.

\begin{figure}[t]
\centering
\includegraphics[width=.5\textwidth]{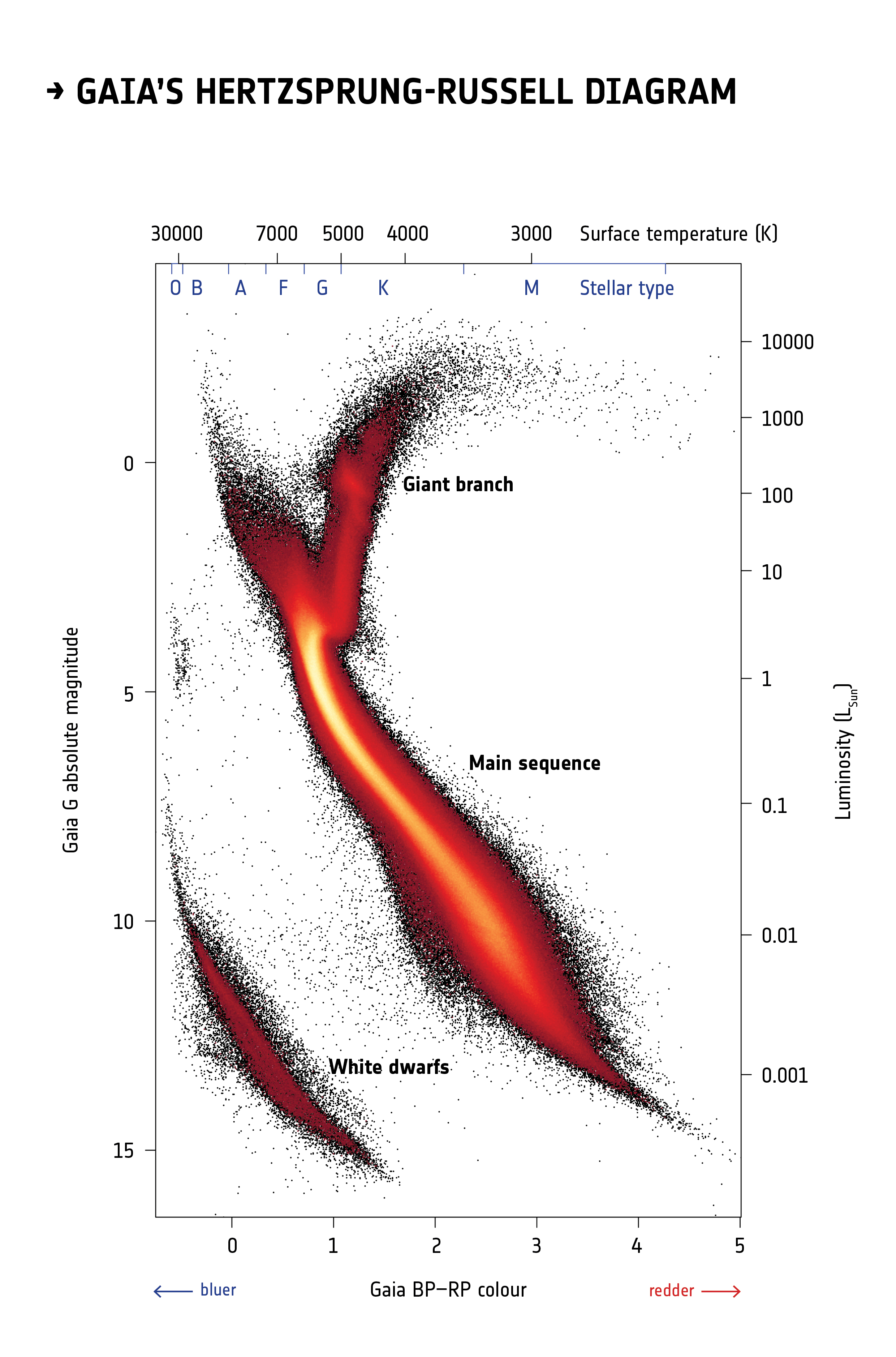}
\caption{Hertzsprung-Russell diagram of all stars identified by the Gaia survey, in Data Release 2, shown as a function of stellar temperature versus luminosity. The main sequence, giant star branch, and white dwarf populations have been labeled. Created by ESA/Gaia/DPAC.}
\label{fig1}
\end{figure}

\section{Radial velocity observations: the first discoveries}

\hspace*{10mm} In the 1970s, the prospect of definitively detecting planets orbiting other stars via radial velocity observations became possible for the first time \citep{campbell1979}. In the decade that followed, searches for planetary signals in nearby, bright stars rapidly expanded. Given the intrinsic relative brightness of giant stars, precision radial velocity studies were soon being focused specifically on giant stars in order to better characterize their variability, and long-period radial velocity observations were observed and published for three K giant stars two years before the unambiguous discovery of a Jupiter-mass planet orbiting a Sunlike star \citep{hatzes1993}. More than a decade later, at least two of these radial velocity oscillation signals were revealed to be due to orbiting planets \citep{hatzes2006, hatzes2015}, making the detection of planets orbiting evolved stars even older than the detection of planets around main sequence stars. These signals were observed in the stars $\alpha$ Tauri and $\beta$ Geminorum, better known as Aldebaran and Pollux, the brightest star in the Taurus constellation and second brightest star in the Gemini constellation, and remain the brightest known planet-hosting stars identified to date.

\hspace*{10mm} Shortly after the first discovery of a planet orbiting a Sunlike star, it became clear that planets could be found on short-period orbits around a wide range of stellar types, triggering searches for planets via the radial velocity technique around hundreds of FGKM stars. A few giant stars were intentionally included in these searches, and a number of these stars were later discovered to be more evolved than initially thought. This resulted in the first unambiguous detection of a planetary-mass object around a post-main sequence star, HD 177830 b \citep{vogt2000}. This object was found to have a minimum mass of approximately 1.5 times the mass of Jupiter ($\approx$ 1.5 M$_\mathrm{J}$), and an orbital period of $\approx$410 days. Its host star was known to be a subgiant, but its evolutionary state more closely resembled that of a main sequence star than a red giant, making the variability of the star easier to characterize and less likely to mask a planetary signal. At the time of detection, the radial velocity variability of giant stars was known to be significantly higher than that of main sequence stars, and thus despite the observed long-period variability in giant stars, they were not widely targeted for planet detection as the larger intrinsic radial velocity variations of giant stars could more easily conceal the signal of an extrasolar planet.

\hspace*{10mm} This changed in the early 2000s, when planet searches with radial velocities also aimed to incorporate astrometric constraints to help confirm the maximum masses of companions detected via the radial velocity method. This triggered a focused search for radial velocity variability among nearby, bright K giants, as their intrinsic brightness allowed them to be characterized astrometrically at much larger distances than was possible for other types of stars. In 2002, the first unambiguous planet candidate detection was made around an evolved star, iota Draconis \citep[][]{frink2002}. This planet candidate was observed to be on a long-period ($\sim$500 day), highly eccentric ($e$ = 0.7) orbit (see Figure 2). Its high eccentricity and larger semiamplitude of variation ($\sim$150 m s$^{-1}$) made the substellar companion origin easier to distinguish from stellar activity, and the non-detection of any astrometric motion of the star confirmed that the maximum mass of the companion had to be below 45 Jupiter masses, ensuring that the companion was not a star.

\begin{figure}[t]
\centering
\includegraphics[width=\textwidth]{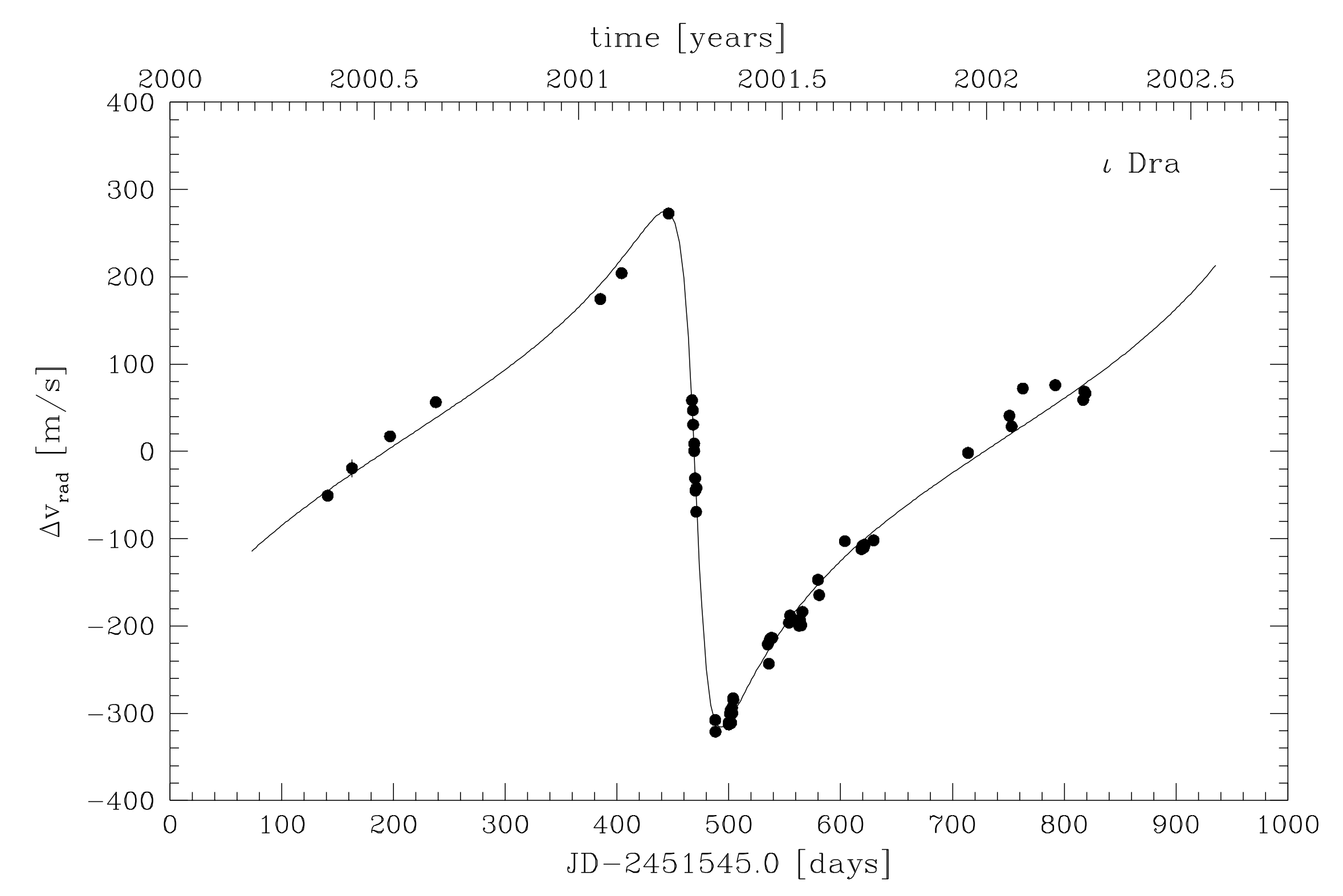}
\caption{Radial velocity signal of iota Dra b, the first planetary mass object detected around a definitively red giant star. Taken from \citet{frink2002}. }
\label{fig1}
\end{figure}

\hspace*{10mm} Although these discoveries confirmed the existence of planets around post-main sequence stars, their large orbital periods were taken as evidence that shorter period planets were not able to survive around such rapidly evolving stars. The detection of the first hot Jupiter orbiting a post-main sequence star did not occur until 2005, at which point a Jupiter-mass planet was found orbiting the subgiant star HD 88133 on a 3.4-day period \citep{fischer2005}. Shortly thereafter, a number of other hot Jupiters were detected around subgiant stars, but none were found orbiting true red giant stars, leading to the belief that planets do not exist at short periods around evolved stars \citep{sato2008}. In order to support this, orbital dynamics theory was developed to explain the observed dearth of planets on short-period orbits around evolved stars \citep{villaver2009}. This dynamical theory suggested that close-in planets ($a$ $<$ 0.5 au) inevitably spiral into their host stars as the host stars increase in size, while planets on wider separations ($a$ $>$ 0.5 au) are flung out to large ($a$ $\sim$ 1.5-5 au) separations by post-main sequence evolution. In addition, this orbital dynamics theory investigated the effect of stellar mass on planet occurrence around evolved stars, as the red giants observed in our own Galaxy tend to span an intermediate range of stellar masses (1-2 M$_\odot$). The ability to use red giant host stars to explore planet occurrence as a function of stellar mass led to a new era in the study of planets orbiting evolved stars.

\subsection{``Retired" A star surveys: key to determining planet occurrence as a function of stellar mass}

\hspace*{10mm} The search for extrasolar planets rapidly intensified after the detection of a planet orbiting a Sun-like star. While planets were relatively easy to detect via the radial velocity method around FGKM stars, the rapid rotation of more massive OBA main sequence stars results in much broader spectral line profiles, making it much more difficult to detect spectral shifts due to an unseen companion. However, as these intermediate-mass stars evolve off the main sequence, they cool, their radii grow and their rotation slows, making their spectral lines much more narrow, thus allowing radial velocity detection of planets. Thus, studying evolved stars made it possible to detect planets around intermediate-mass stars, and the study of ``retired A stars" to determine planet occurrence as a function of stellar mass was born.

\hspace*{10mm} By 2010, there were sufficient detections of planets orbiting evolved stars to allow population level studies of these systems in context of all other known planetary systems. In addition, a hot Jupiter was found around a massive subgiant star, HD 102956 b, which was claimed to be the most massive star to harbor a hot Jupiter with a stellar mass greater than 1.68 $\pm$0.11 M$_\odot$ \citep{johnson2010b}. Using this and similar detections of exoplanet systems, a correlation between planet occurrence and stellar mass was claimed to continue from low-mass stars through intermediate-mass stars, but relied almost entirely on evolved stars to represent the population of intermediate-mass stellar planet hosts \cite[][see Figure 3]{johnson2010a}. Our ability to determine masses of red giant stars was then heavily dependent on the accuracy of stellar models, as the evolutionary tracks of different stellar masses on the giant branch are much more closely packed in color-magnitude space than on the main sequence. Different models could result in significantly lower masses inferred for these stars \citep[e.g.,][]{lloyd2011}. This warranted a new approach to comparing the planet populations of main sequence and evolved stars. 

\hspace*{10mm} Later searches of planets orbiting evolved stars considered hundreds of star in their sample, allowing relative mass measurement within the evolved population of stars without the consideration of main sequence planet hosts \citep[e.g.,][]{wittenmyer2011}. These studies similarly recover a trend of increase in occurrence with stellar mass, with a peak in occurrence around 2 M$_\odot$ with a drop in planet occurrence at larger masses. These systems also show an increase in planet occurrence with stellar metallicity \citep{reffert2015}. More recently, radial velocity surveys have reproduced these results with respect to stellar mass, and suggest a peak in planet occurrence at orbital periods near 700 days, with planet occurrence decreasing at larger separations, in similar proportions as to those seen around main sequence stars \citep[][see Figure 4]{wolthoff2022}. 

%However, planet detection from other approaches appeared poised to change this, as the first direct imaging detections of exoplanets were made around young A stars \citep{marois2008}.

\begin{figure}[t]
\centering
\includegraphics[width=\textwidth]{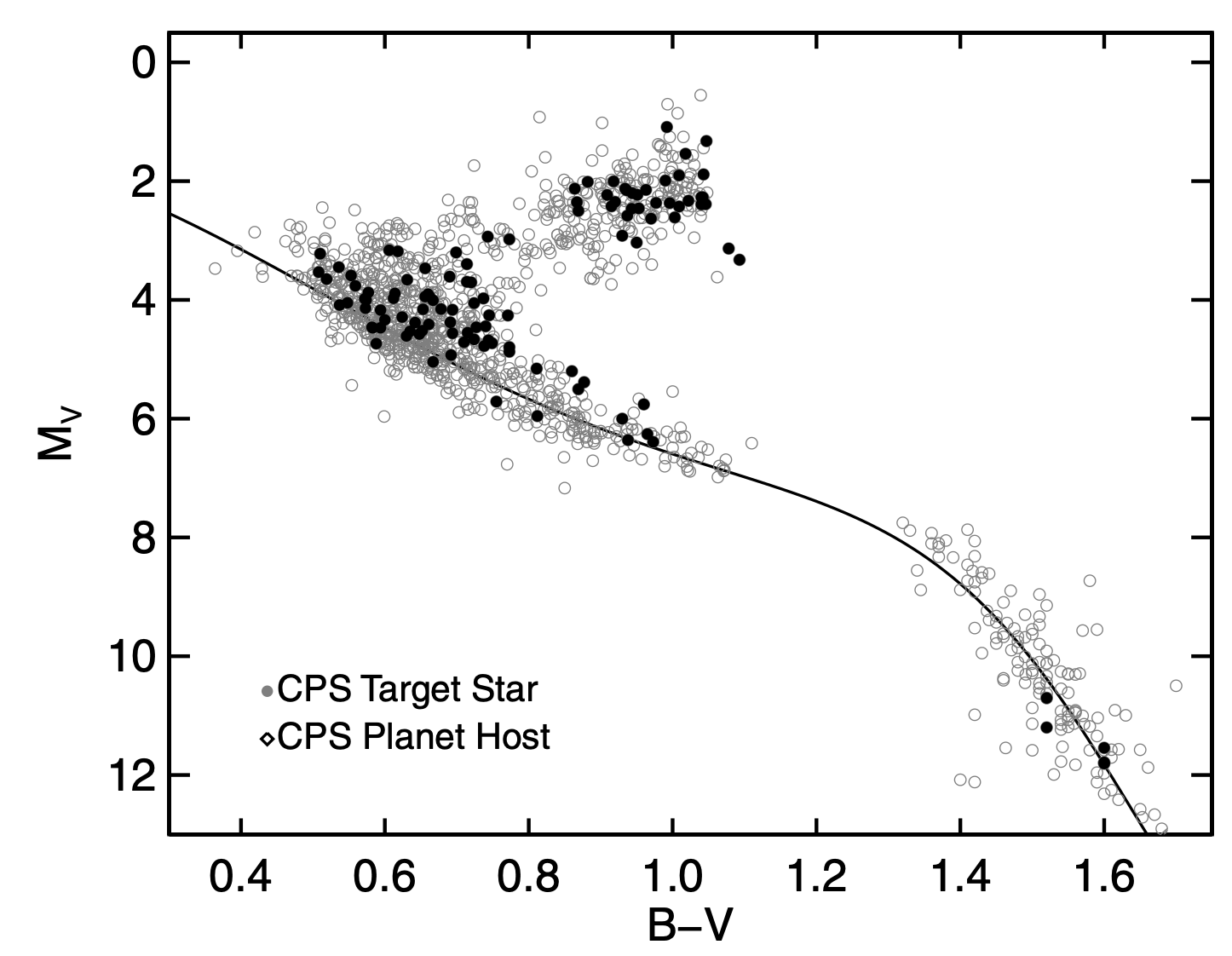}
\caption{Absolute magnitude in the V bandpass versus the difference in stellar B and V magnitude. This example of a Hertzsprung-Russell diagram shows the planet detections that had been made by the California Planet Search (CPS) team by 2010. Giant stars consisted of a significant fraction of all surveyed planet hosts. This population was used to draw conclusions about planet occurrence as a function of stellar mass, yet the high-luminosity and thus high-mass end of the population were essentially entirely evolved systems, blurring the distinction between stellar mass-driven and evolution-driven effects on the planet population. Modified from \citet{johnson2010a}.}
\label{fig3}
\end{figure}

\begin{figure}[t]
\centering
\includegraphics[width=\textwidth]{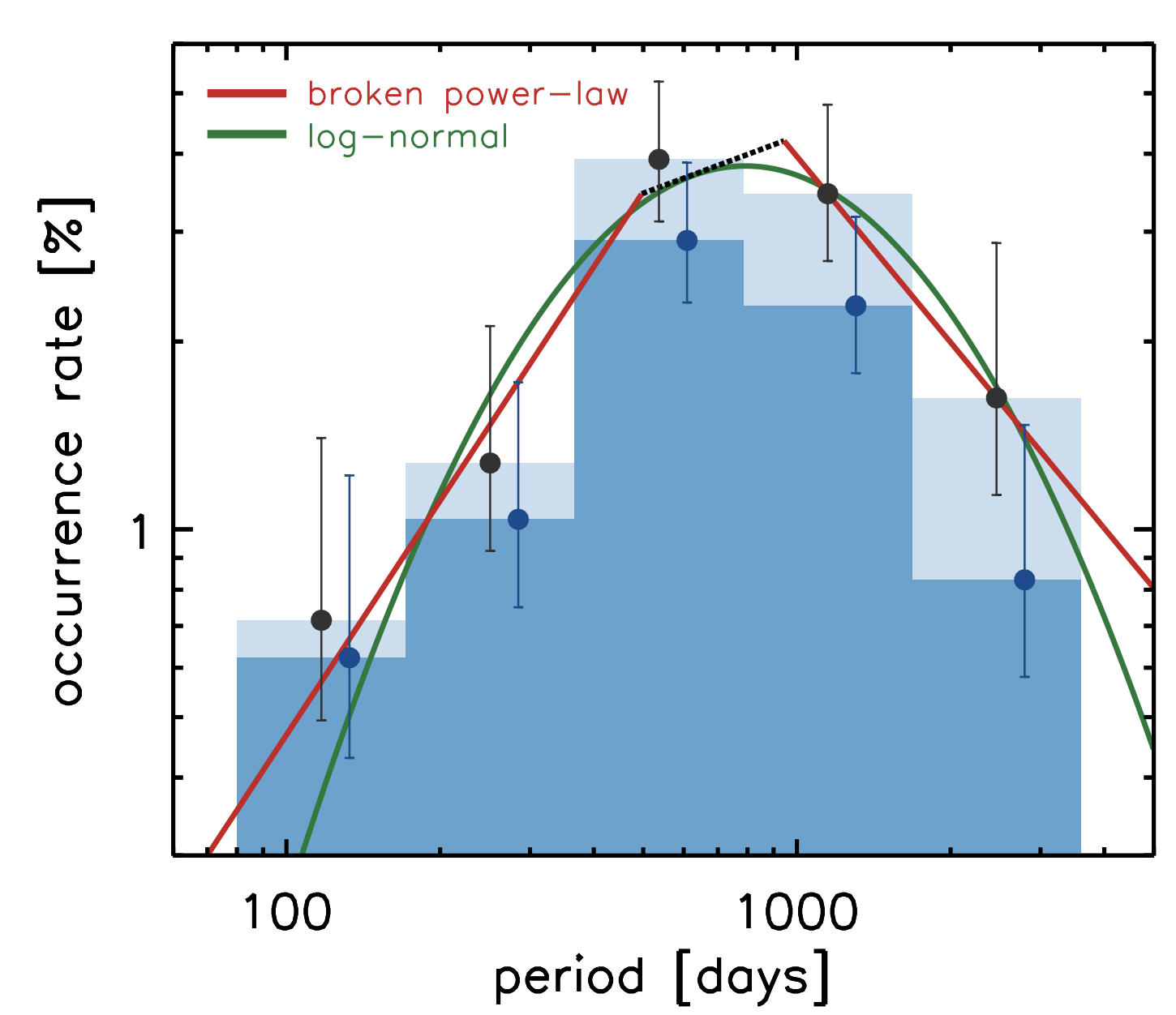}
\caption{Planet occurrence as a function of orbital period, as determined by \citet{wolthoff2022}. A turnover in planet occurrence is seen between 1-3 au, similar to what has been seen in main sequence radial velocity determined occurrence rates.}
\label{fig4}
\end{figure}

\section{Transit surveys}\label{chap1:sec2}

\hspace*{10mm} While the radial velocity planet community was debating the true nature of planets detected around evolved stars, new opportunities for planet detection were beginning in space. The NASA \emph{Kepler} mission was launched in 2009 to detect planets via the transit method, where the planet passing in front of the star in its orbit is recorded as a periodic dip in the brightness of the star. \emph{Kepler} quickly proved to be the most successful planet-hunting mission to date, detecting thousands of new planets over a span of a few years. However, similar to previous transiting planet surveys, \emph{Kepler} specifically avoided targeting giant stars, as their large radii and intrinsic photometric variability made planet transits more difficult to detect. Nevertheless, a significant number of red giant stars were serendipitously observed by \emph{Kepler}, due to large uncertainties in its initial stellar input catalog. A few of these red giant stars happened to host transiting planets. 

\subsection{Breakthroughs by accident: the Kepler era}

\hspace*{10mm} Though relatively few in number, the planets discovered around evolved stars by the \emph{Kepler} mission were essential to improving our understanding of planet populations overall. The discovery of Kepler-91 b, a hot Jupiter orbiting a star whose radius is more than 6 times larger than our Sun \citep{lillobox2013}, revealed that planets could be found on short period orbits around red giant stars. This system also demonstrated that the oscillations of red giant stars could be modeled using Gaussian process estimation and removed from transit light curves in order to obtain more precise transit parameters \citep{barclay2015}. Simultaneously with the discovery of Kepler-91 b, the Kepler-56 system revealed that multiplanet systems could survive until early red giant branch evolution \citep{huber2013}. Such a system could not be confirmed by radial velocity measurements alone. These discoveries opened the door to future searches for planets transiting evolved stars, and developed the current notion that short-period planets orbiting evolved stars may be equally as common as planets orbiting main sequence stars at short period but simply harder to observe.

\hspace*{10mm} The detection of these systems was also aided by the detection of stellar oscillations in the light curves of their host stars. The analysis of these oscillations, known as asteroseismology, allowed precise determination of stellar masses and radii using a method independent of stellar models. This updated approach to determining stellar properties resulted in a significant reduction in uncertainty of radii and masses of red giant stars, often constrained to within 5\% uncertainty or less. This has also allowed determination of stellar spin-orbit obliquity, which revealed a misalignment in the Kepler-56 system \citep{huber2013}. Later light curve models incorporated both planet transit features as well as the ability to model stellar oscillations and granulation as simple harmonic oscillators operating as Gaussian processes \citep[e.g.,][]{grunblatt2017}. The ability to conduct transit searches and asteroseismology simultaneously has allowed new depth in the investigations of evolved systems, allowing new insight into how planetary evolution is influenced by stellar evolution in the era of exoplanet surveys \citep{grunblatt2019}.

\subsection{Targeted surveys: \emph{K2} and \emph{TESS}}

\hspace*{10mm} After the failure of 2 reaction wheels on board the \emph{Kepler} spacecraft, the \emph{K2} Mission was started in 2013. Instead of targeting one specific field, the \emph{K2} Mission observed several different fields along the ecliptic for $\sim$90 days at a time. The Mission also allowed specific target requests from the community, allowing evolved stars to be specifically targeted for transiting planet observations for the first time. Such a focused survey allowed the comparison of main sequence analogs to evolved systems. By providing radius constraints for several short-period planets orbiting red giants, the effect of the post-main sequence increase in irradiation could be directly observed \citep{lopez2016}. Asteroseismology of these same targets also helped to characterize these unique systems more precisely. Detection of two very similar, short-period, inflated hot Jupiters orbiting evolved stars suggested that the late-stage increase in stellar irradiation was directly contributing to planet inflation in these systems, as similar mass planets orbiting similar mass stars at similar orbital periods were found to be significantly smaller \citep{grunblatt2016,grunblatt2017}. This was supported by future occurrence studies, which revealed that hot Jupiters orbiting evolved stars appear significantly larger than those orbiting main sequence stars, on average \citep{grunblatt2019}. This survey also revealed that the overall occurrence of hot Jupiters was statistically indistinguishable between main sequence and evolved stars, suggesting that hot Jupiters orbiting evolved stars were both inflated and not as rapidly doomed relative to their main sequence analogs as was previously believed (see Figure 5).

\begin{figure}[t]
\centering
\includegraphics[width=\textwidth]{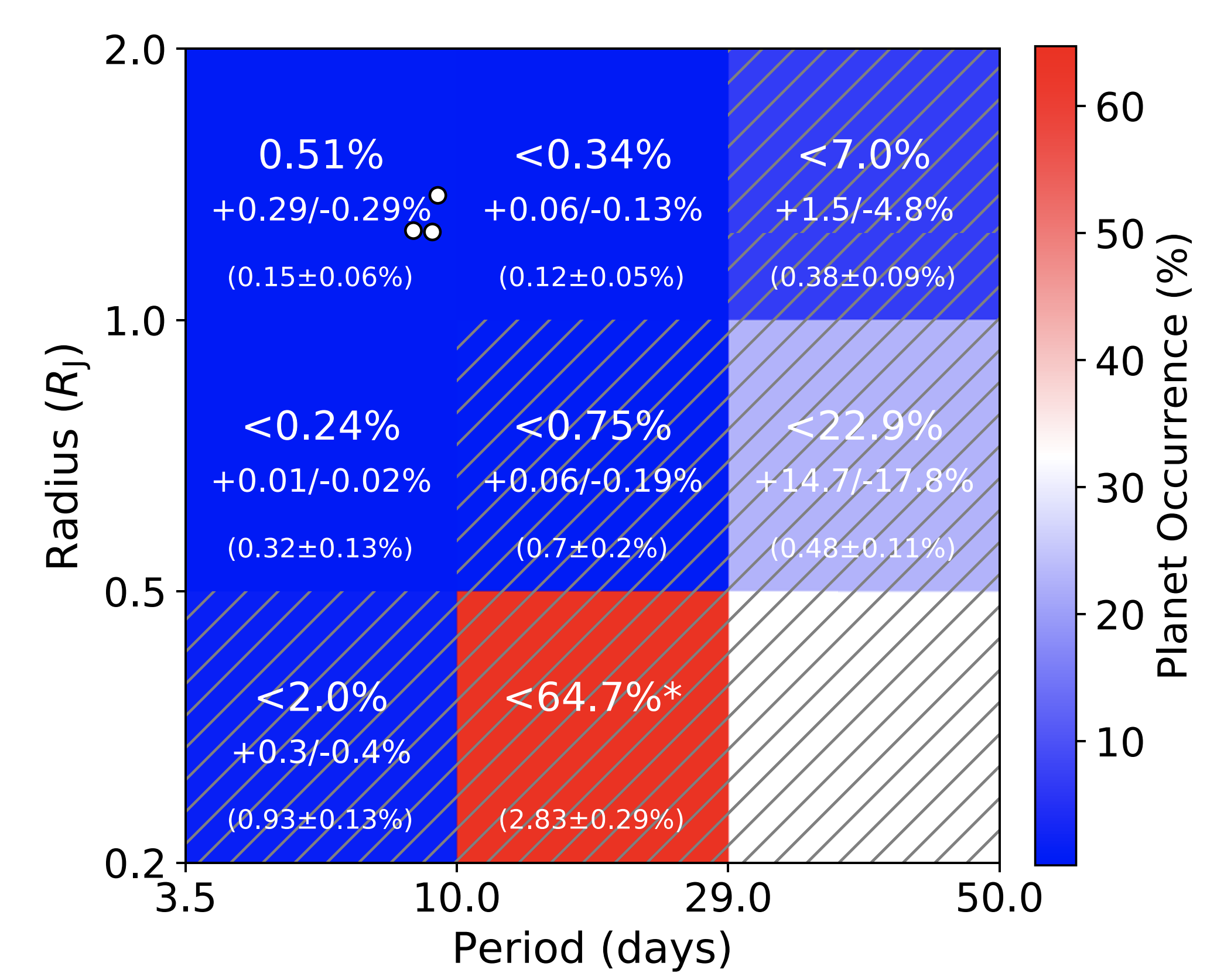}
\caption{Planet occurrence around 3.5–8 R$_\odot$ stars observed by \emph{K2}, as a
function of orbital period and radius. In those bins where no planets were
found, upper limits were calculated for planet occurrence. Hatched cells
designate where injection/recovery completeness is below 50\% for this
sample, and the asterisk indicates where uncertainties on completeness were
too large to be reliably estimated. Main-sequence occurrence rates from
\citep{howard2012} are shown in parentheses at the bottom of each bin.
Planets detected by \citet{grunblatt2019} are shown by the white circles. For planets
with radii larger than Jupiter at orbital periods less than 10 days, a
consistent yet tentatively higher number of planets orbiting our sample of
LLRGB stars than main-sequence stars is found. For all regions of parameter space
where planets were not found, the upper limits of planet occurrence calculated
by this survey are in agreement with the main-sequence occurrence rates
reported by \citep{howard2012}. Taken from \citet{grunblatt2019}.}
\label{fig5}
\end{figure}

\begin{figure}[t]
\centering
\includegraphics[width=\textwidth]{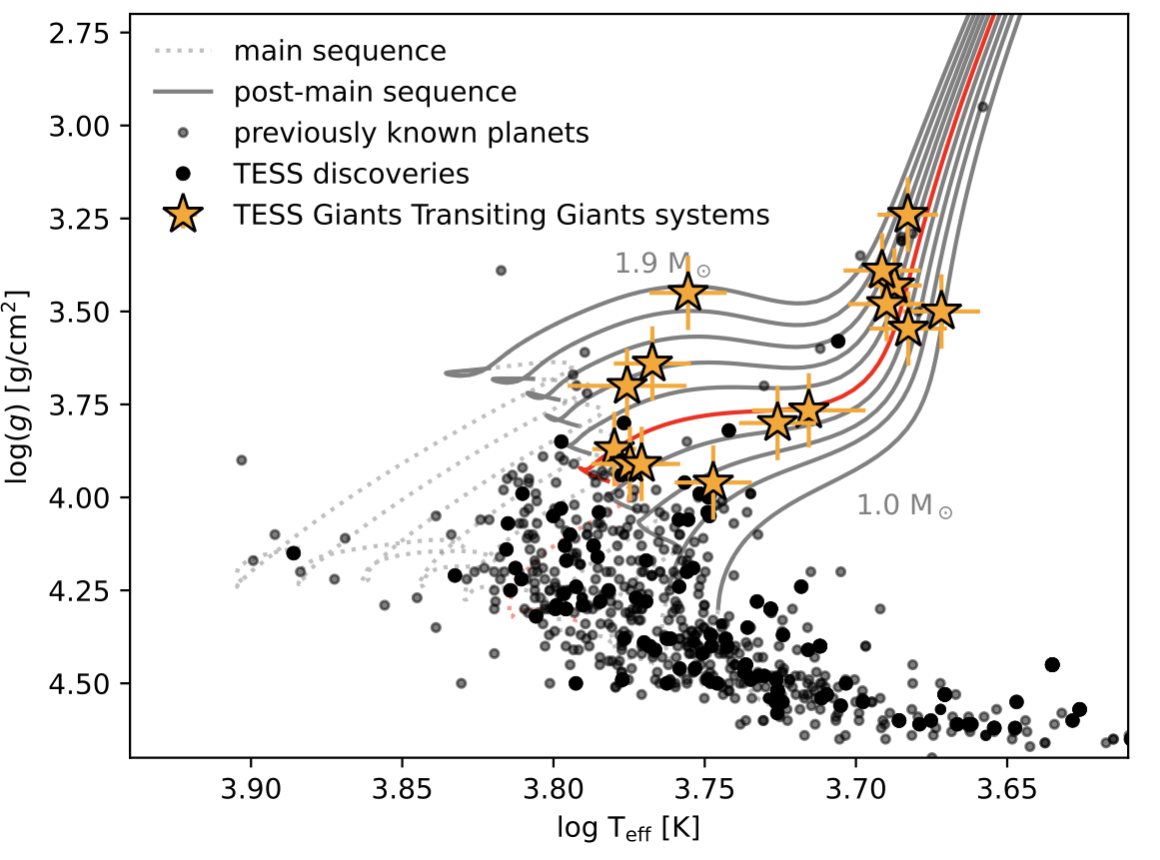}
\caption{Planets discovered by the \emph{TESS} Giants Transiting Giants survey, shown as a function of stellar effective temperature and surface gravity, both in log units. MIST stellar evolutionary tracks for [Fe/H]= 0.25 dex stars between 1.0 and 2.0 solar masses are shown in gray and red dotted and solid lines. Stellar parameters for known planet discoveries are shown as gray dots, where TESS discoveries are shown as larger, black dots. Stars hosting planets discovered by the TESS Giants Transiting Giants Survey are shown as orange stars.}
\label{fig6}
\end{figure}

\hspace*{10mm} After \emph{K2}, the all-sky survey led by \emph{TESS}, or the Transiting Exoplanet Survey Satellite, made it possible to detect an order of magnitude more planets transiting evolved stars than was possible with any other transit survey to date. This survey has revealed a wide range of radii and masses for hot and warm Jupiters orbiting evolved stars, with over three times the number of transiting planets found around evolved stars by the \emph{TESS} Mission than by any previous missions. This has largely been driven by the \emph{TESS} Giants Transiting Giants project, which has confirmed more than 10 planets transiting evolved stars with \emph{TESS} so far \cite[e.g.,][see Figure \ref{fig6}]{grunblatt2022}. In addition, the all-sky survey nature of \emph{TESS} has allowed the detection of a number of transiting hot planets around main sequence A stars \citep[e.g.,][]{zhou2019}, thus allowing direct comparison of transiting planet populations at different evolutionary states at a wide range of orbital periods. The advent of \emph{TESS} along with galactic kinematics from the \emph{Gaia} Mission also allowed for the first systematic search of planets transiting halo stars. The stars targeted in this search were almost exclusively giants because of their brightnesses relative to other halo stars. While no planets were found to be transiting halo stars, the constraints provided suggest that a survey of roughly ten times the number of halo stars has the potential to detect a transiting planet in our Galactic halo \citep{yoshida2022}.

\subsection{Future surveys}

\paragraph{PLATO}

\hspace*{10mm} The \emph{PLATO} mission aims to detect planetary transits and oscillations of stars. Launching in 2026, the initial plan for this ESA space mission is to stare at one field near the southern ecliptic pole for over 2 years. With a field of view significantly larger than \emph{Kepler} and a pixel scale and limiting magnitude fainter than \emph{TESS}, \emph{PLATO} is expected to discover tens to hundreds of planets transiting evolved stars, for which asteroseismology to determine precise stellar parameters should be possible in the vast majority of cases. However, as the planned observations of \emph{PLATO} are not yet finalized or available to the public, it is difficult to say exactly how planet candidates orbiting evolved stars will be vetted by the mission, and thus its impact on our current understanding of planets orbiting evolved stars is largely unknown.

\paragraph{Roman}

\hspace*{10mm} The future NASA \emph{Roman} Mission, and particularly the Galactic Bulge Time Domain survey of the \emph{Roman} mission, is predicted to find an order of magnitude more planets than has ever been found before, including thousands of hot Jupiters transiting evolved stars \citep{wilson2023}. These systems will be detected at kiloparsec distances, with over 100 systems predicted to be detected around red giant stars beyond 10 kiloparsecs from the Sun, on the opposite side of our Galaxy. Thus a survey of evolved hot Jupiters with Roman has the capacity to reveal changes in evolved planet occurrence as a function of the radius of our Galaxy, linking exoplanet studies to Galactic archaeology in a way that cannot be achieved with earlier missions. Successful detection of distant planets with the Galactic Bulge Time Domain Survey with Roman could lead to characterization of even more distant or fainter planet hosts with \emph{JWST}, which could reveal transiting planets in unique Galactic substructures such as globular clusters or satellite galaxies.

%\begin{figure}[t]
%\centering
%\includegraphics[width=.4\textwidth]{blankfig}
%\caption{A conservation relationship can also be written for electric charge, but in mesoscale modeling, electromagnetic effects are not considered to be dynamically or thermodynamically important on the model-resolved mesoscale. They are certainly important on  cloud and precipitation microphysics, and can therefore affect mesoscale
%and larger processes, but this would need to be included through.}
%\label{fig1}
%\end{figure}

\section{Evolved Planet Populations}\label{chap1:sec3}

\hspace*{10mm} With over three decades of history, planets orbiting evolved stars have now been detected and confirmed through multiple observation methods, and we can begin to probe the underlying planet distribution while controlling for some observational biases. Though no rocky planets have yet been detected around evolved stars, larger planets that are detectable, such as Neptune- and Jupiter-sized planets, have been found at a range of separations from their host stars, with a wide range of masses and orbital eccentricities. The distributions of these properties are suggestive of multiple evolution pathways which occur on different timescales that can be more clearly distinguished among the evolved transiting planet population. In addition, evidence for rocky material orbiting white dwarfs suggests that rocky planets must exist around post-main sequence stars as well, and thus we discuss some of the scenarios which provide evidence for this rocky planetary material and pathways by which it may be processed over time. We discuss some of the most well-studied planet subpopulations around evolved stars, and the new astrophysics that can be learned by studying these planet populations. 

\subsection{Hot Jupiters}

\hspace*{10mm} Since the Nobel Prize-winning discovery of a Jupiter-sized planet orbiting a Sunlike star at a separation smaller than 0.1 au, astronomers have debated the potential origin of such extreme planetary systems. In general, the proposed origins of hot Jupiter systems can be split into three classes: {\it in situ} formation, disk migration, and tidal migration. Each of these formation mechanisms operate on different timescales and require different protoplanetary disk architectures, and the resultant planetary system architecture and longevity is closely related to the initial formation mechanism. The unique evolutionary state of planets orbiting post-main sequence stars makes these systems particularly valuable to distinguish between different potential hot Jupiter formation and evolution mechanisms. Thus, astronomers are currently striving to understand the properties of this unique population of planets and place them in context of their main sequence analogs to understand how planetary systems evolve and change over time.

\paragraph{Overcoming early biases}

\hspace*{10mm} Though planets orbiting evolved stars were arguably detected before planets around main sequence stars, the first detection of a hot Jupiter orbiting an evolved star was almost 15 years after the first detection of a hot Jupiter orbiting a main sequence star. This led early researchers to conclude that hot Jupiters are rare around evolved stars. However, more recent studies have shown that hot Jupiters are equally common around main sequence and evolved stars, if orbital separations out to $\sim$0.1 au are considered \citep{grunblatt2019, temmink2023}. At orbital periods less than 3 days, or orbital separations of less than 0.05 au, planets around evolved stars are significantly less common, due to the intense tidal effects at these small separations from large stars. This difference also suggests that hot Jupiters seen around main sequence and post-main sequence stars may have arrived at their current states through different evolutionary pathways. 

\hspace*{10mm} Why have radial velocity surveys historically been so bad at detecting hot Jupiters orbiting evolved stars? One reason is simply geometrical--many of the evolved stars observed in radial velocity surveys were too large to be able to host planets within 0.1 au, as their radii were often 0.05 au ($\sim$10 R$_\odot$) or larger. Another was the high amount of stellar variability, which made the detection of weak signals difficult, but also distorted sinusoidal planetary signals on hour- to day-long timescales, close enough to the planetary orbital period to make hot Jupiter orbit detection significantly more difficult. In addition, the slightly longer period distribution of hot Jupiters orbiting evolved stars meant that significantly longer baselines were often necessary to detect evolved hot Jupiters than main sequence hot Jupiters, and the semiamplitude of the radial velocity measurements were lower while the stellar signal baseline activity is higher, making hot Jupiter detection more difficult from the ground. 

\paragraph{Orbital processes: laboratories}

\hspace*{10mm} While hot Jupiters transiting giant stars are relatively rare outcomes of planetary system evolution, they hold valuable insight into the evolution of all planetary systems. Given the larger orbital momenta and smaller separations involved, there are a number of aspects of star-planet interaction which manifest themselves in the properties of evolved hot Jupiter systems but are invisible in earlier-stage or smaller-planet systems. Here we discuss a subset of the most well-known planetary system orbital evolutionary features that can be much more easily probed in evolved planetary systems.

\subparagraph{Circularization and Inspiral: a population perspective}

\hspace*{10mm} The initial assumed dearth of planets at short periods orbiting evolved stars was believed to be due to the combined effect of orbit circularization and inspiral triggered by the acceleration of the evolution of the host star on the subgiant and red giant branch of stellar evolution (see chapter 'Giant branch systems: dynamical and radiative evolution' for more details). Though the original dearth of planets predicted by these evolutionary processes has been refuted, this circularization evolutionary pathway for post-main sequence planetary systems remains, and should result in the destruction of many main sequence hot Jupiter systems during post-main sequence evolution. Thus, the population of hot Jupiters around evolved stars likely have different origins than the analogous population orbiting main sequence stars.

\hspace*{10mm} The first hot Jupiter to be discovered transiting an evolved star was Kepler-91 b \citep{lillobox2013}, a $\sim$0.8 M$_\mathrm{J}$ planet orbiting a red giant star every 6.25 days. Given the short period of the planet and the relatively evolved state of the host star (R$_*$ = 6.38 R$_\odot$), this planet was quickly recognized to be near the end of its life, with the star growing to the orbit of the planet in 55 Myr, and inspiral processes likely leading to the demise of the planet before that date. In addition, the orbit of the planet appears to be mildly eccentric ($e \sim$ 0.05). Given the rapidly changing stellar radius, this mild eccentricity suggests that the planet orbit is actively circularizing. During the post-main sequence phase of stellar evolution, circularization and inspiral of planetary orbits occur simultaneously, where neither process can complete alone. Thus, this suggests that Kepler-91 b is doomed to be engulfed by its host star in the next tens of millions of years, and was likely not a hot Jupiter during the main sequence lifetime of Kepler-91. 

\hspace*{10mm} Discovery of additional post-main sequence hot Jupiters built off of this discovery to probe the sensitivity of orbital circularization and inspiral to stellar and planet parameters. Five years after the initial discovery of Kepler-91 b, a handful of hot Jupiters were confirmed orbiting evolved stars \citep[e.g.,][]{grunblatt2016}. Interestingly, several of these hot Jupiters were observed to be on non-circular orbits, suggestive of late-stage orbit circularization. Eventually, population-level studies revealed that relative to main sequence stars, hot Jupiters orbiting evolved stars had significantly more eccentric orbits \citep{grunblatt2023a}. This further cemented the idea that planets orbiting evolved stars have different origins than the populations of planets orbiting main sequence stars, and the timescale for the formation of these evolved hot Jupiter systems was governed by different processes than main sequence hot Jupiters \citep{villaver2009}.

\begin{figure}[t]
\centering
\includegraphics[width=\textwidth]{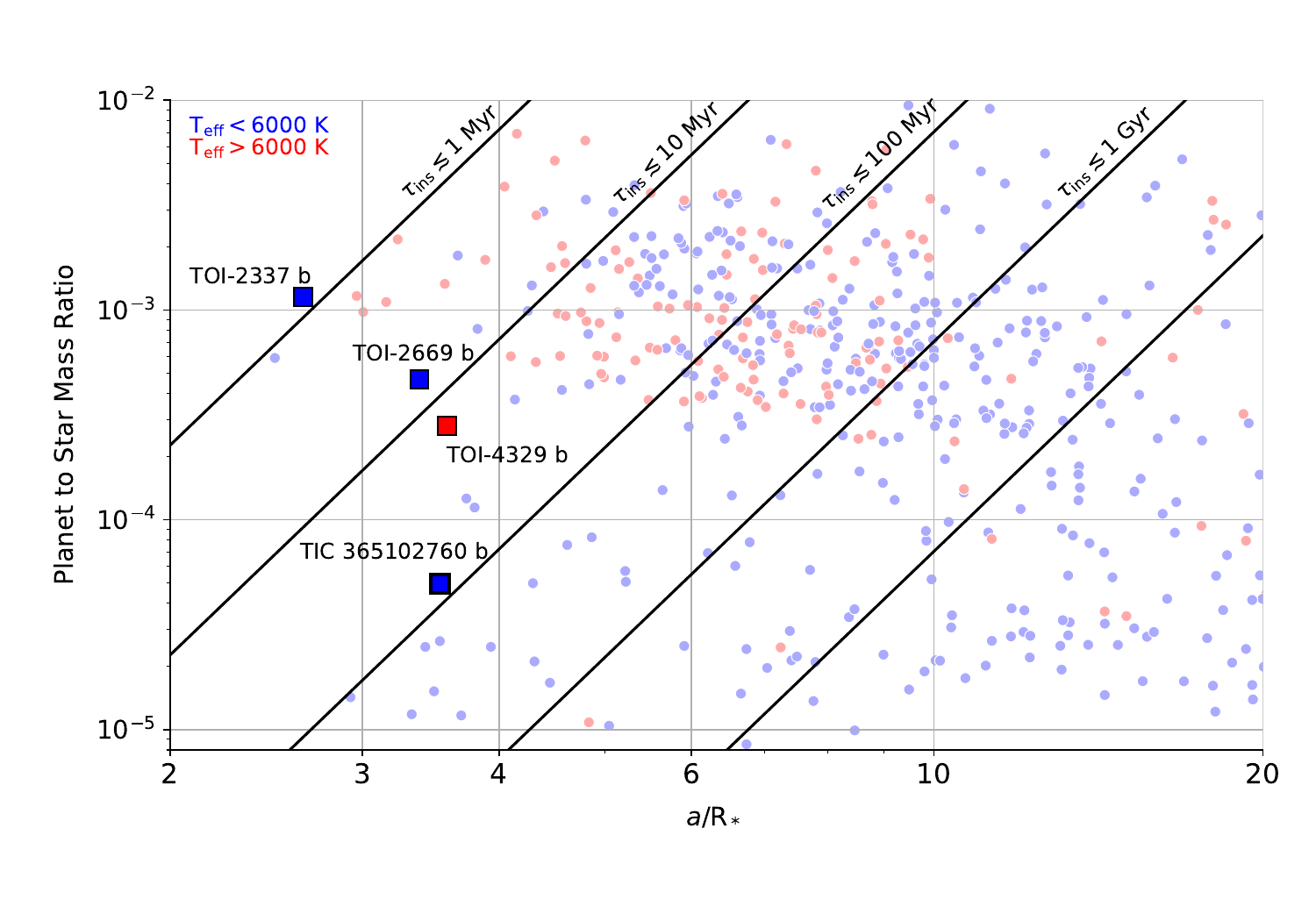}
\caption{Star to planet mass ratio, versus orbital separations scaled by the stellar radius, for confirmed exoplanet systems. Orbital decay timescales decrease toward the upper left of this plot, where black diagonals correspond to theorized rates of orbital decay, where the leftmost line corresponds to a decay timescale of 10$^6$ years, and each following line increases by a factor of 10. Blue points have stellar effective temperatures $<$6000 K as reported by the NASA Exoplanet Archive, while red points represent planets around hotter stars. The planets confirmed by \citet{grunblatt2022} and \citet{grunblatt2024} are shown as squares with black outlines, and are populating relatively sparse regions of parameter space on this plot that correspond to rapid orbital decay. In particular, TIC 365102760 b may be experiencing the fastest rate of orbital decay of any planet with a mass less than 1/10,000th that of their host star, and TOI-2337b may be experiencing the fastest rate of orbital decay of any planet known to date. Modified from \citet{grunblatt2022}.}
\label{fig7}
\end{figure}

\hspace*{10mm} Theoretical predictions of planetary inspiral timescales were not testable until the first direct detection of planetary inspiral in the hot Jupiter system WASP-12b \citep{maciejewski2016}. The evolutionary state of this system has been debated, but the inspiral of the hot Jupiter suggests that the system has evolved off of the main sequence, as orbital decay is expected to rapidly accelerate during post-main sequence evolution even though the dominant mechanism for orbital decay is still unclear. Orbital decay has now been confirmed in at least one additional hot Jupiter system, which also happens to be in a post-main sequence evolutionary state \citep{vissapragada2022}. Preliminary evidence for orbital decay has been claimed in several other systems, and verification or refutation of these claims will reveal the role of stellar evolutionary state in triggering hot Jupiter inspiral. In addition, since the discovery of Kepler-91 b, hot Jupiters have continued to be found on shorter and shorter orbits around evolved stars, with multiple planets now confirmed to be transiting evolved stars at orbital periods less than 3 days \citep{grunblatt2022}. Given the incredibly small star-planet separations in these systems, the predicted inspiral timescales of these evolved systems are shorter than those in the vast majority of other exoplanet systems. One recently detected evolved hot Jupiter, TOI-2337 b, is predicted to inspiral into its host star in less than one million years (see Figure \ref{fig7}). Continued observations of known hot Jupiter systems by the NASA \emph{TESS} Mission are expected to confirm orbital decay in several more systems over the next 5 years, allowing exploration of the dependence of this process on star and planet properties, and constraining the tidal quality factors of evolved host stars observationally for the first time. 

\hspace*{10mm} In addition, the most recent population level searches for hot Jupiters transiting evolved stars suggest that there may be two sub-populations of hot Jupiters orbiting evolved stars, which have arrived at their current architecture through different pathways, as discussed above. The confirmation and characterization of several evolved hot Jupiter systems identified by \emph{TESS} suggest that hot Jupiters orbiting younger, hotter, less-evolved post-main sequence stars are more likely to be more massive and on circular orbits, while hot Jupiters orbiting older, cooler, more evolved stars are more likely to have moderately eccentric ($e \sim 0.15$) orbits and show evidence of additional companions in the system \citep{chontos2024}. Constraining the absolute occurrence of these types of systems among evolved stars will help to provide insight into the different formation and evolution pathways of planetary systems, and how planetary system evolution is correlated with star and planet properties.

\subparagraph{Spin-orbit obliquity}

\hspace*{10mm} Over the last 15 years, it has become clear that the origin and evolution of hot Jupiter systems, as well as evolved systems, can be understood by measuring the angle or obliquity between the orbital plane of the hot Jupiter and spin axis of the host star. In transiting hot Jupiter systems, this projected spin-orbit obliquity can be measured via the Rossiter-McLaughlin effect, where the radial velocity signal of the star is measured as the planet transits the star, and the change in measured stellar radial velocity over the course of the transit reveals the relative obliquity of the planet's orbit to the star's spin axis. The earliest measurements of the Rossiter-McLaughlin effect suggested that most hot Jupiters were well aligned with their host stars, but subsequent measurements of hot Jupiters orbiting higher temperature ($T_\mathrm{eff}$ $>$6250 K) stars with radiative, as opposed to convective, outer envelopes revealed hot systems host a much wider range of obliquities between the stellar rotation and planetary orbit than cooler systems \citep{hebrard2008}. 

\begin{figure}[t]
\centering
\includegraphics[width=\textwidth]{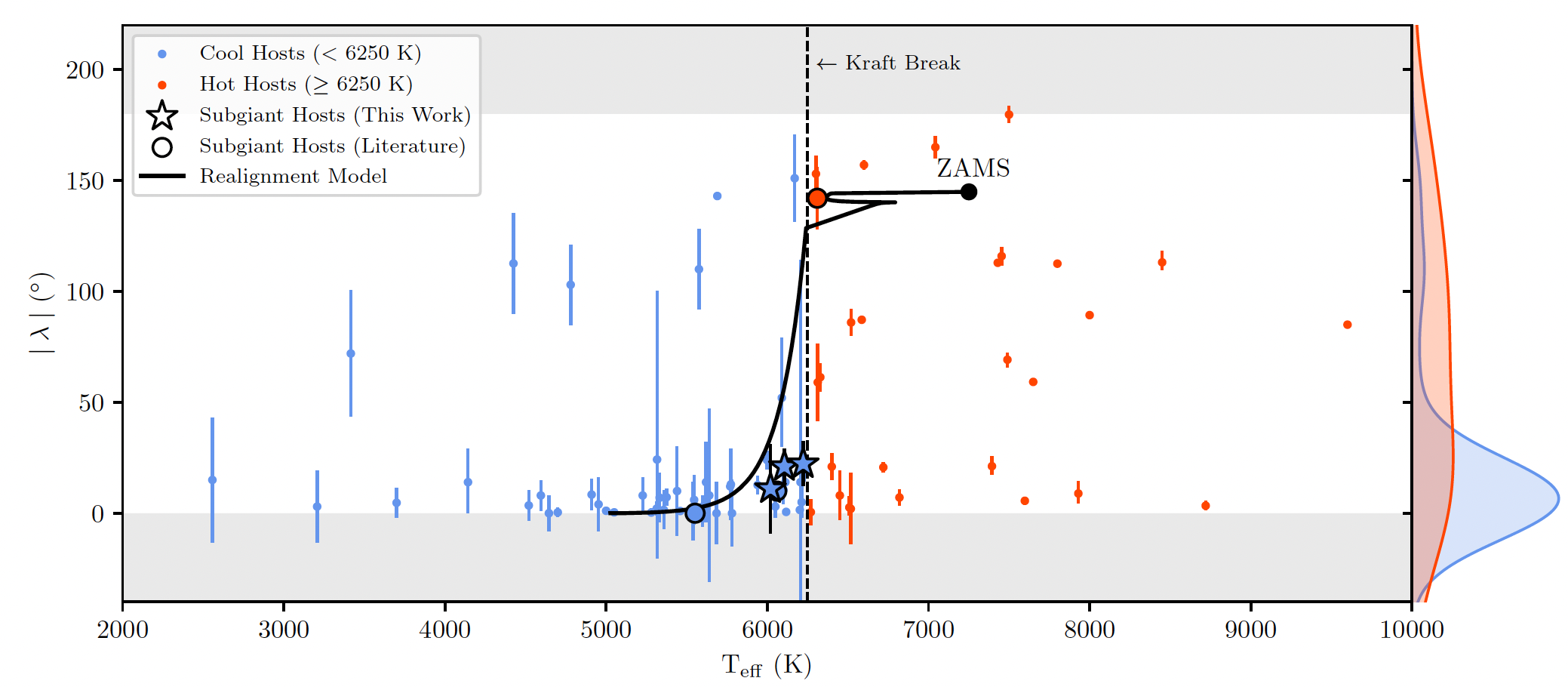}
\caption{Distribution of hot Jupiter obliquities as a function of stellar effective temperature. Blue points indicate cool ($<$6250 K) hosts, and red points indicate hot ($>$ 6250 K) hosts. The vertical dashed black line marks the delineation between these populations. Subgiant hosts are marked by large circles with black outlines, and all other points are main sequence hosts. The distributions on the far right of the plot show the kernel densities of each sample, corresponding to color. The black line shows a model for the post-main sequence realignment of a hot Jupiter, shown in $\lambda$ versus $T_\mathrm{eff}$. This model shows a possible pathway from the hot, misaligned population to the observed cool, well-aligned population. Provided by N. Saunders (priv. comm.).}
\label{fig11}
\end{figure}

\hspace*{10mm} Evolved systems have proved to be particularly valuable for understanding the spin-orbit obliquity distributions of planetary systems. The majority of evolved hot Jupiter systems known today are systems whose host stars would have been hotter than 6250 K on the main sequence but have cooled to temperatures closer to 5000 K on the subgiant and red giant branches. The first measurement of stellar spin-orbit misalignment in a multiplanet system was made in an evolved star system, suggesting a different formation for this system than other planetary systems orbiting cooler stars \citep{huber2013b}. Subsequent Rossiter-McLaughlin measurements of hot Jupiters orbiting evolved stars suggest that these systems seem to be misaligned around hotter host stars, but then appear well aligned around cooler host stars with convective envelopes (N. Saunders, priv. comm., see Figure \ref{fig11}), despite the widespread misalignment of the precursors to these systems. This provides further evidence that hot Jupiter systems around evolved stars likely have different origins than those around main sequence stars, and also is suggestive of an upper limit of tens of millions of years for alignment to occur in hot Jupiter systems orbiting stars cooler than 6250 K. However, significantly more measurements of spin-orbit obliquity in late-stage hot Jupiter systems must be made before these timescales and evolutionary pathways can be well defined.

\paragraph{Atmospheric processes: laboratories}

\hspace*{10mm} Closely tied to planetary orbital evolution, the evolution of planetary atmospheres can also be probed by hot Jupiters transiting evolved stars in unique ways. As the incident flux on a planet changes, this affects the heating on the planet, which is widely believed to have strong implications for the atmospheric circulation and structure of planets. If the irradiation on the planet is strong enough, it can also result in the stripping of a planetary atmosphere over time, further shaping the observed distribution of exoplanetary systems. Here we explore the observational constraints on the heating and stripping of hot Jupiter atmospheres around post-main sequence stars.

\subparagraph{Re-inflated planets}

\hspace*{10mm} After the discovery of the first known hot Jupiter, planetary structure theorists began to predict how the atmosphere of a hot Jupiter would be influenced by the intense irradiation of their host stars. Within a year of the discovery of 51 Pegasi b, it was suggested that the structure of hot Jupiters must be influenced by their host star irradiation, resulting in an outer radiative zone where heat was transferred into the interior of these planets \citep{guillot1996}. The first measurement of a hot Jupiter radius, made possible by the first measurement of a planet transiting its host star, was observed to be larger than standard planetary structure models would allow, and was in good agreement with predictions of planetary atmospheric inflation driven by external irradiation \citep{charbonneau2000}. However, less than a year after this measurement was published, it was suggested that the large radius of this planet was not driven by external radiation, but rather that the internal heat from formation of this planet was never released, due to the heat flow in the environment of a hot Jupiter, and thus allowed the planet to maintain a larger radius than expected \citep{burrows2000}. In addition, it was also suggested that the inflation of these planets may also be driven primarily by tidal dissipation in the planets as they circularize their orbits over time \citep{bodenheimer2001}, though the detection of inflated planets on stable orbits around old stars suggested that tidal dissipation did not operate over long enough timescales to explain the observed planet population. The debate between whether hot Jupiters were inflated directly by stellar irradiation or indirectly by maintaining their initial heat from formation remained unresolved for decades.

\hspace*{10mm} With the advent of transiting hot Jupiter discoveries around evolved stars, new tests to the distribution of planet parameters became possible for the first time. The detection of hundreds of hot Jupiter systems revealed an incident flux threshold for planet inflation, near 150 times the flux on Earth, below which no inflated planets were observed. Theorists soon realized that a comparison among a subset of hot Jupiter systems could distinguish between planet inflation models that required direct heat deposition from the star into the planet versus models which suggested the planets never lost their heat from formation. By comparing planets at orbits where the planets were below the inflation threshold when their host star was on the main sequence, but above the inflation threshold at evolved stages, the origin of planet inflation could be probed directly: if the population of planets orbiting evolved stars were significantly larger than their main sequence counterparts, this suggested that the planets had become 're-inflated' by the increase in irradiation of their host stars as they evolved off of the main sequence, but if the radii of both the main sequence and evolved planets were similar, this suggested that the planets were simply experiencing delayed cooling after their initial formation, and were not influenced directly by the evolution of their host star \citep{lopez2016}. 

\begin{figure}[t]
\centering
\includegraphics[width=\textwidth]{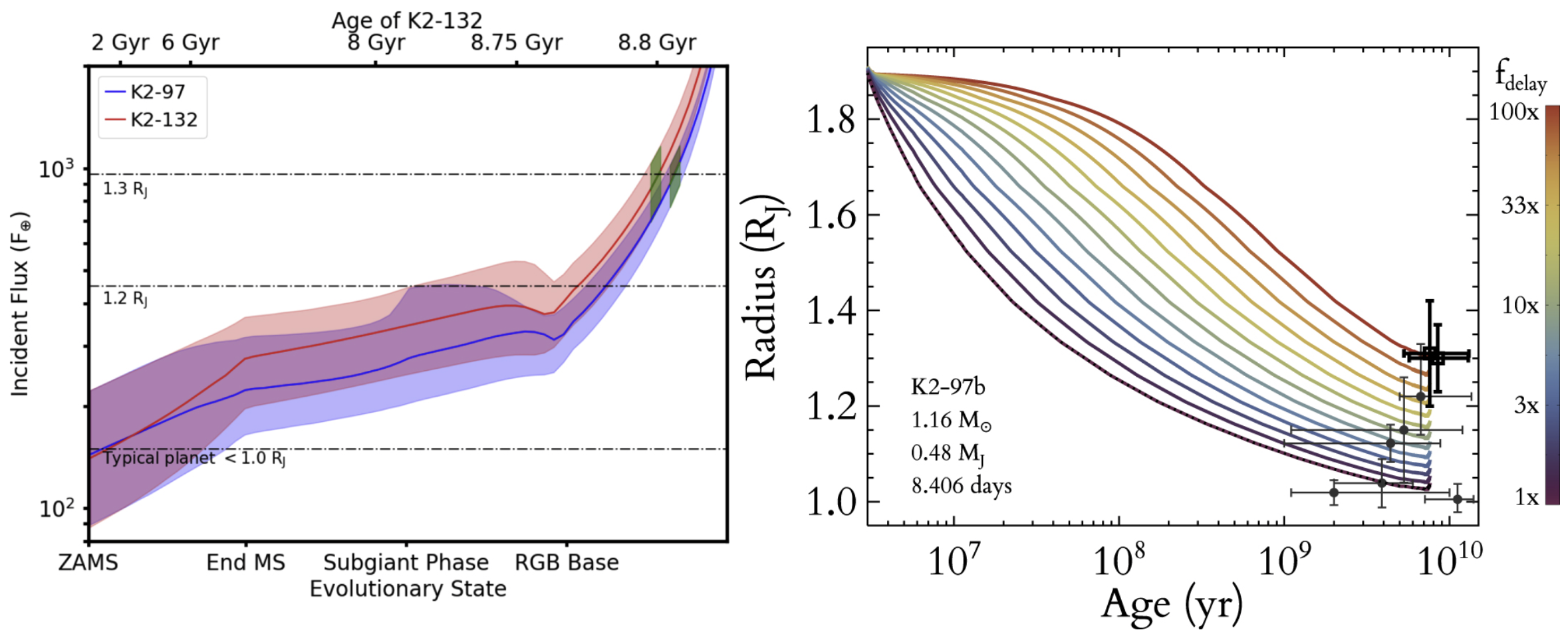}
\caption{ {\it Left:} Incident flux as a function of evolutionary state for K2-97b and K2-132b. The top axis shows representative ages for the best-fit stellar parameters of K2-132. The current incident flux on the planets is denoted in green. The solid blue and red lines and shaded areas show the median and 1-$\sigma$ confidence interval considering uncertainties in stellar mass and metallicity. The black dashed lines correspond to the median incident fluxes for known populations of hot gas giant planets of different radii. The observed radii of 1.3 R$_\mathrm{J}$ for K2-97 b and K2-132 b imply they are typical for hot Jupiters receiving their current irradiation, not their main sequence irradiation, suggesting that external heating is responsible for the re-inflation of these planets (NASA Exoplanet Archive, 2017 September 14).  {\it Right:} Planetary radius as a function of time for K2-97b and K2-132b (bold), as well other similar mass planets with similar main-sequence fluxes orbiting main-sequence stars. Colored tracks represent scenarios where planets begin at an initial radius of 1.85 R$_\mathrm{J}$ and then contract according to the Kelvin–Helmholtz timescale delayed by the factor given by the color of the track. All main-sequence planets seem to lie on tracks that would favor different delayed cooling factors than the post-main-sequence planets shown here. Taken from \citet{grunblatt2017}.}
\label{fig8}
\end{figure}

\hspace*{10mm} Once hot and warm Jupiters were detected at slightly longer (8-15 d) orbital periods around evolved stars, this comparison could be made directly. The discovery of two very similar hot Jupiters transiting red giant branch stars revealed that both planets were inflated, despite having incident fluxes near or even below the inflation threshold during their main sequence lifetimes \citep{grunblatt2016, grunblatt2017}. With radii close to 1.3 R$\mathrm{Jup}$, these planets were typical in size for gaseous planets receiving their current incident flux, but are significantly larger than similar mass planets receiving their main sequence incident flux (see Figure \ref{fig8}, left). In addition, using a delayed cooling model to describe the population of planets with similar masses and orbital periods required a span in delayed cooling rate of over two orders of magnitude (see Figure \ref{fig8}, right), while the entire population could also be explained by a much smaller range of re-inflation models, where only one heating efficiency could explain the appearance of the whole population. When additional inflated Jupiters with similar masses were found around evolved stars more recently, their appearances could also be explained by an identical heating efficiency for re-inflation.

\hspace*{10mm} However, there are still many aspects of planet re-inflation that are not understood, such as the dependence of re-inflation on planet properties. Planet mass is known to be an important factor in the efficiency of planet re-inflation, and there may be a planet mass above which planet inflation simply does not operate. \emph{TESS} discoveries of somewhat more massive hot Jupiters orbiting post-main sequence stars have revealed planets at or above the mass of Jupiter receiving even higher incident fluxes than the known re-inflated planets that do not appear to be inflated at all, despite the recent increase in irradiation on these planets driven by the evolution of their host stars \citep[e.g.,][]{saunders2022}. This suggests that the sensitivity of a hot Jupiter to inflation is inversely proportional to the planet's mass. Additionally, lower mass planets on similarly short orbits have also been identified to be transiting evolved stars. Although the inflation rates of these planets are harder to measure due to uncertainty in the size expected for such low-mass, low-density planets, the existence of such planets is certainly easier to explain by invoking planet re-inflation at late times. In addition, studying these lower-mass planets around evolved stars helps to constrain another dominant force in sculpting the appearance of these planets: atmospheric mass loss.

\subparagraph{Atmospheric mass loss (in evolved systems)}

\hspace*{10mm} The highly irradiated environment of hot Jupiters makes these planets susceptible not only to atmospheric inflation, but also the loss of atmospheric material. Ultraviolet and X-ray radiation can dominate the evolution of these planets' upper atmospheres, where the gas temperature nears the atmospheric escape temperature, and thus the absorption of a high energy photon can result in the escape of atmospheric material from the planet's gravitational influence. The evaporation of planet atmospheres was first detected via the observation of an extended planetary atmosphere in the H$\alpha$ bandpass relative to its appearance in white light \citep{vidalmadjar2004}. This suggested that material may be escaping from the atmosphere of the planet. Additional detections of extended atmospheres in other stellar lines and excess absorption continuing after the end of the planet transit have confirmed this loss of atmospheric material in exoplanetary systems.

\begin{figure}[t!]%{r}{0.5\textwidth}
\centering
\includegraphics[width=0.85\linewidth]{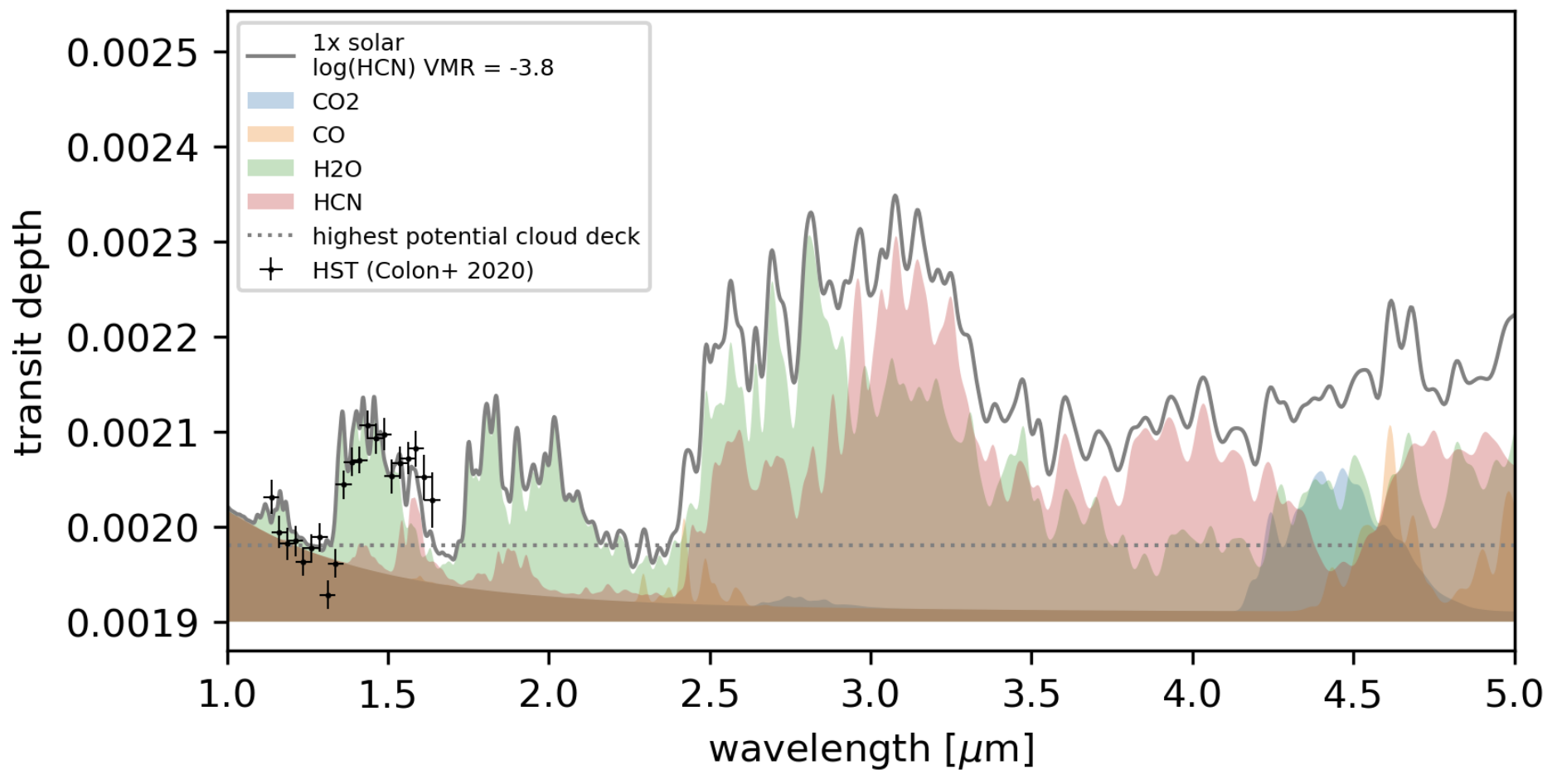}
\includegraphics[width=0.85\linewidth]{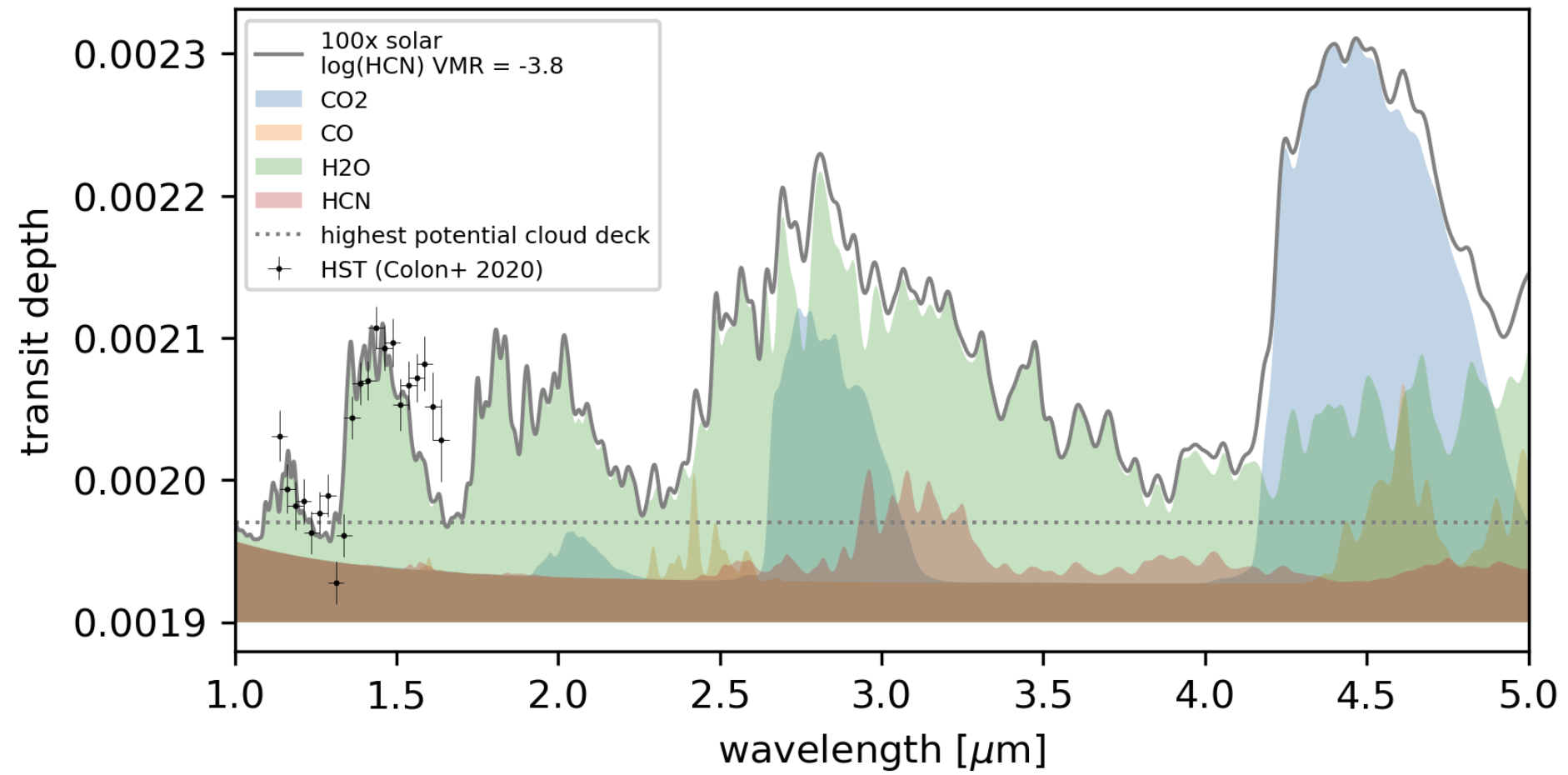}
%\vspace{-0.5cm}
\caption{Hubble Space Telescope transit observations of KELT-11 b, a low-density planet orbiting a subgiant star. Measurements are taken from \citet{colon2020}, overlaid on atmospheres simulated with equilibrium chemical species as well as nonequilibrium volume mixing ratios (VMRs) of hydrogen cyanide (HCN), as suggested by \citet{colon2020}, assuming a planet metallicity equal to that of the Sun (top), or 100x that of the Sun (bottom), in the wavelength ranges where the James Webb Space Telescope is sensitive. Opacity contributions of CO$_2$, CO, H$_2$O, HCN and clouds are shown individually in both models by the blue, orange, green, and red filled regions, and the dotted gray line, respectively. The larger absorption features at 3-5 microns require much higher cloud decks to be muted by clouds. Nonequilibrium HCN will be dominate the spectrum in the 3-5 $\mu$m wavelength region in a low (1x solar) metallicity atmosphere, but still produces features detectable with high precision in a high (100x solar) metallicity atmosphere, with comparable or better detection of equilibrium species like CO, CO$_2$ and H$_2$O in all cases.}
\label{fig9}
\end{figure}

\hspace*{10mm} As some of the most mature exoplanet systems, hot Jupiters orbiting evolved stars are predicted to have experienced some of the largest fractional amounts of atmospheric mass loss of any gaseous planets. In addition to planet inflation mechanisms, atmospheric mass loss can also contribute to decreasing the density of these planets, as well as changing the overall composition of the planet's atmosphere. Observations of a low-density hot Jupiter orbiting an evolved host, KELT-11, with the Hubble Space Telescope have revealed intriguing deviations from the expected transmission spectrum of the planetary atmosphere, which are currently not well understood, but match similar deviations seen in other planets transiting post-main sequence stars \citep[e.g.,][]{colon2020}. These deviations are believed to be related to disequilibrium chemistry in the planetary atmosphere, which is suggestive of significant vertical mixing in the planetary atmosphere, another process necessary for planet re-inflation. Future observations with the James Webb Space Telescope can reveal a much larger wavelength range of planetary atmospheric transmission, providing a much better constraint on the composition of these planet atmospheres (see Figure \ref{fig9}). In addition, for the largest planets, such constraints on the atmospheric composition of evolved hot Jupiters may be possible using ground-based facilities. These observed differences may be even larger in the lowest mass planets, where atmospheric mass loss has the largest potential to modify the composition of planetary atmospheres by stripping lighter elements such as hydrogen and helium while the planets retain heavier compounds in their atmospheres. 

\subsection{Longer period planets}

\hspace*{10mm} In addition to the hot Jupiter population, the population of longer period giant planets orbiting evolved stars also appear to have features which are distinctly different from the main sequence planet population. These features can reveal aspects of planetary system evolution at wider separations, testing the stability and longevity of planets on Earthlike orbits as their host stars evolve.

\begin{figure}[t]
\centering
\includegraphics[width=.8\textwidth]{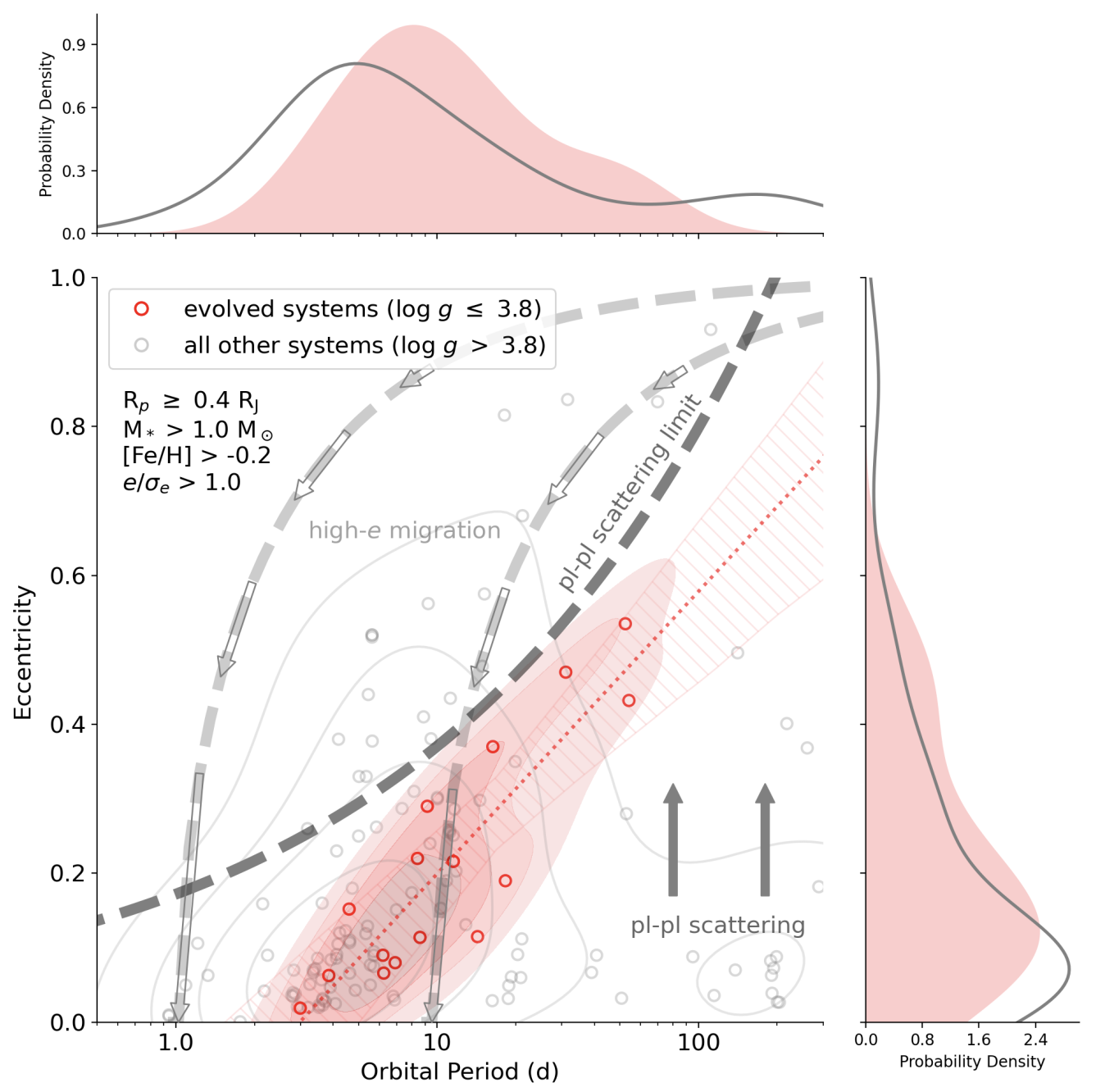}
\caption{Eccentricity as a function of orbital period for planets in systems where R$_p$ $\geq$ 0.4 R$_\mathrm{J}$, orbiting stars where M$_*$ $>$
1.05 M$_\odot$ with metallicities $>$ -0.2 dex, with significant eccentricities. Systems transiting evolved (log(g) $\leq$ 3.8) stars have been
shown by the red contours and corresponding red points, while planets transiting main sequence (log($g$) $>$ 3.8) stars are shown
by the gray contours and points. Tracks of constant orbital angular momentum followed during high-eccentricity migration
are shown by the light gray dotted lines and arrows, while the effects of planet-planet scattering and its approximate limit 
are shown by the dark gray arrows and dotted line.
We find that planets transiting evolved stars appear to form a more linear distribution in period-eccentricity space than the rest
of the planet population, possibly sculpted by these migration and scattering processes. Furthermore, we find that the orbital
eccentricity of evolved systems can be approximated well by a linear regression to the logarithm of the orbital period, shown by
the red dotted line and hatched region corresponding to a 95\% confidence interval. A similar linear correlation is significantly
weaker for the overall planet population. Taken from \citet{grunblatt2023a}.}
\label{fig10}
\end{figure}

\hspace*{10mm} Most notably, the confirmation of several longer-period ($>$10 d) planets transiting evolved stars revealed that these planets appear to be exclusively on eccentric orbits. Moreover, longer period evolved planets orbits appear to be more eccentric than shorter period planet orbits, revealing a log-linear relationship between orbital eccentricity and period for planets transiting evolved stars, which is not seen in equivalent main sequence systems \citep[][see Figure \ref{fig10}]{grunblatt2023a}. Considering this along with the predicted circularization and inspiral of these systems, this suggests that planets on eccentric orbits around evolved stars circularize and inspiral as their stars undergo post-main sequence evolution, potentially destabilizing planets on interior orbit in the process. Meanwhile, planets on orbits outside of these inspiraling planets will likely move to larger orbits to preserve angular momentum, and will likely remain on more circular orbits than planets closer in, as has been observed for planets detected around evolved stars with orbits larger than 1 au.

\hspace*{10mm} Both of these events will increase the likelihood of planet-planet scattering in these systems, as the relative distances between planets change. This will have the largest effect closer to the host star, where planet-planet scattering events are more likely, which can in turn increase the eccentricity of planet orbits, causing the process to repeat itself, triggering a positive feedback loop in such systems at smaller separations. The detection of longer period companions to a number of eccentric, evolved hot Jupiters further supports this evidence for late-stage planetary system restructuring that is divergent at different orbital separations from the star \citep{chontos2024}. Additional constraints on the existence of long period companions to evolved hot Jupiters will help to confirm or refute the currently existing theories of late-stage planetary system evolution.

\hspace*{10mm} The effect of stellar multiplicity on planetary system is also not particularly well understood. The dearth of planets known in late-stage multi-star systems may suggest that the increased dynamical complexity of such systems curtails the longevity of planetary systems, but may also be related to observational biases which prevent the systematic study of such systems. The detection of brown dwarfs transiting evolved stars, as well as the detection of close-in planets orbiting old stars in binary systems, suggests that evolved planetary systems in binaries can survive \citep{campante2015, page2024}. Improvements in simulations that connect stellar evolution to planetary orbital evolution will help to predict the occurrence of such systems.

\subsection{Multiplanet systems}

\hspace*{10mm} Few multiplanet systems have been identified around evolved stars. Currently, only one multi-transiting system with a giant star host is known \citep{huber2013b}. This is likely related to detection biases when searching for planets around these types of stars: multiplanet systems tend to feature planets smaller than Jupiter, yet very few planets smaller than Jupiter are detectable around evolved stars due to the stars' large radii. 

\hspace*{10mm} The one currently known multi-transiting system with an evolved stellar host also displays a number of other features which may provide clues to its uniqueness. The system in question, Kepler-56, shows evidence for at least three substellar components. The two innermost components are transiting planets on 10 and 20 day orbits, respectively, and are among the lowest-mass planets ever detected in evolved planetary systems. These innermost planets are also near a 2:1 orbital resonance, and their orbital plane is clearly misaligned with the rotation axis of the host star. An outer, more massive companion has been detected with an orbital period of $\approx$1000 days. Similar detection of long-period radial velocity variability in hot Jupiter systems may be evidence for a similar evolutionary path between the two classes of evolved systems, but additional detections of multiplanet systems or tighter constraints on their occurrence in post-main sequence systems are needed before the effects of stellar evolution on multiplanet systems can be meaningfully quantified.

\subsection{Evidence for planet engulfment}

\hspace*{10mm} One of the most intriguing recent developments among the population of planet-hosting evolved stars is the prospect of detecting evidence for planet engulfment. Planet engulfment is expected to be a relatively common outcome of late-stage planetary system evolution, due to the high likelihood of planetary inspiral during red giant branch stellar evolution \citep{villaver2009}. The detection of high abundances of lithium or refractory elements in the photospheres of stars may be a sign of planetary engulfment, but may also be a relic of the material which formed the star in the first place, or other poorly understood processes within the stellar interior. Thus, the easiest way to search for long-lasting evidence of such engulfment is to identify relative overabundances of refractory, or rock-forming, elements in individual components of multistar systems. The detection of such refractory element enhancement has been clearly identified in more than one case, but has not yet been clearly distinguished from primordial abundance differences. A high lithium abundance has also been proposed as a tracer of planet engulfment, but interpreting the lithium abundance of evolved stars is generally more difficult outside a narrow range of stellar mass and evolutionary state, as lithium can also be destroyed and/or created in the cores of evolved stars. Similar pollution of stellar photospheres by planet-forming material has been observed for over a century in white dwarf stars, and planet candidates have recently been detected around white dwarf stars but the connection between white dwarf planetary material and refractory overabundances in main sequence stars is not yet clear. 

\hspace*{10mm} In addition to chemical evidence for planet engulfment via spectroscopy, the photometric signature of post-main sequence planet engulfment has been theorized, and one potential photometric signal of a planet engulfment event matching this description has now been reported \citep{de2023}. However, as the theory and observation of photometrically-detected engulfment even is in its infancy, further studies are needed to reveal the occurrence and variance of these events.

\subsection{Future of evolved planet population studies: planets across the Galaxy}

\hspace*{10mm} Though a number of mysteries of the planet population of evolved stars have been solved in the last 10 years, there are still many more questions that remain to be fully answered. Though we now have a better understanding of the role of star-planet interaction in these systems, significantly more systems need to be confirmed to fully explore dependencies on star and planet properties, and understand the connections between main sequence, giant branch, and white dwarf planetary systems. 

\begin{figure}[t]
\centering
\includegraphics[width=\textwidth]{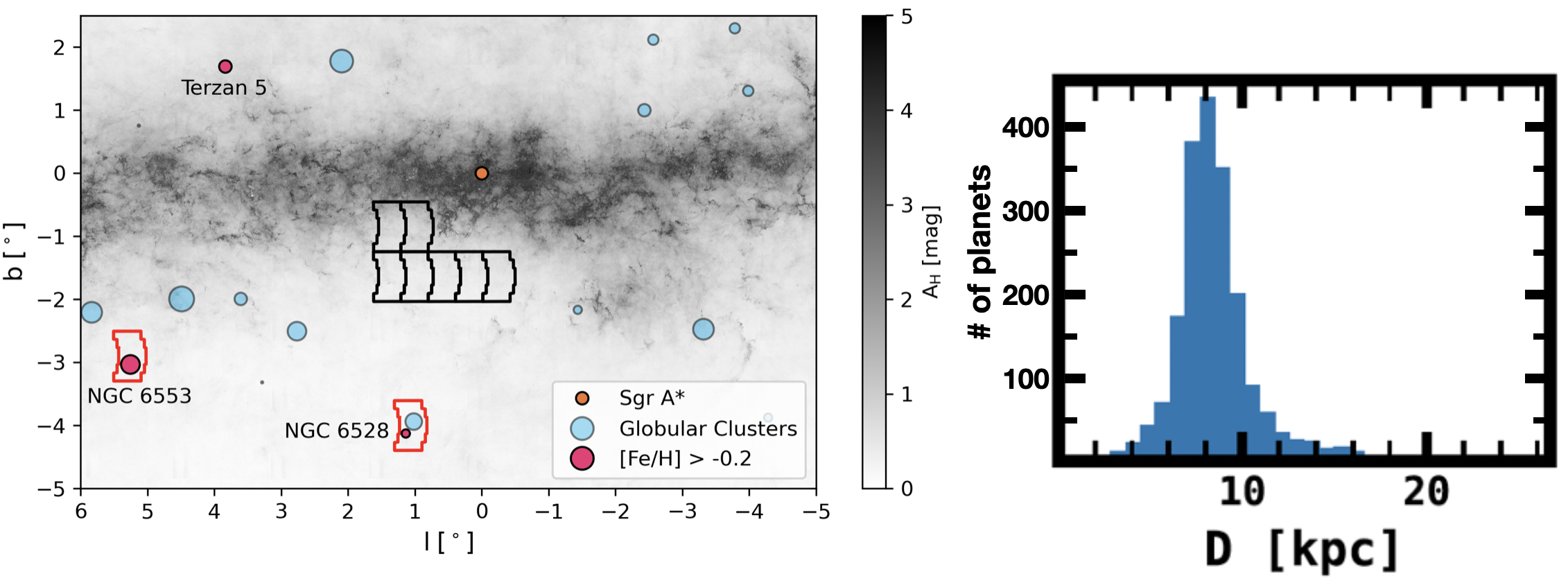}
\caption{ {\it Left:} The footprint of the proposed Galactic Bulge Time Domain Survey, overlaid on a dust map showing magnitudes of distinction in the H band of the Galactic Center. We have designated globular clusters as circles on this plot, where the point size is relative to the angular size of the cluster. Solar-like metallicity globular clusters have been highlighted and labeled. We demonstrate that both NGC 6522 and NGC 6528 can be observed in one Roman pointing only a couple of degrees away (red footprint) from the bulge survey fields. Taken from \citet{grunblatt2023c}. {\it Right:} Predicted distances to \emph{Roman} evolved transiting planet hosts. Planets transiting evolved hosts found by \emph{Roman} will act as beacons for planet populations across the Galaxy, allowing planet occurrence to be estimated in all regions of the Milky Way. Modified from figure provided by R. Wilson (priv. comm.)}
\label{fig12}
\end{figure}

\hspace*{10mm} The NASA \emph{Roman} Space Telescope, scheduled to launch in late 2026, is predicted to detect thousands of hot Jupiters transiting evolved red giant stars as part of its Galactic Bulge Time Domain Survey \citep{wilson2023}. These systems, unlike main sequence systems, are expected to be detectable at all distances within the Milky Way, with over 100 planets expected to be detected transiting red giant stars at distances larger than 10 kpc (see Figure \ref{fig12}). Furthermore, \emph{Roman} can also be used to survey globular clusters of stars for transiting planets using a similar survey strategy, measuring the occurrence of planets in varied birth environments across the Milky Way. This rapid increase in the number of evolved hot Jupiter systems, along with the wide spatial distribution of these systems, means these systems will allow for the first studies of planet occurrence and potential habitability of different regions of the Galaxy, as well as resolving additional issues regarding the formation and evolution of planetary systems. In addition, this strategy can be used to target stellar populations which are believed or known to have extragalactic origins, and may reveal the first planets formed in other galaxies which can be independently verified through multiple methods. Soon, red giant planetary systems will act as beacons for understanding planet occurrence and variance across our Galaxy.

\section{Conclusion}

\hspace*{10mm} Planets have been detected and confirmed around evolved stars for as long as they have been detected around main sequence stars. The demographics of planets orbiting evolved stars remains more poorly understood than that of main sequence stars, but recent discoveries of planets transiting giant stars, particularly hot Jupiters, are closing that gap. Planets orbiting evolved stars support evidence for planet re-inflation and change in planetary orbital eccentricity, semimajor axis and spin-orbit obliquity at late evolutionary stages. Future studies of planets around evolved stars will reveal planets on larger separations than are currently possible, and will let us study planet populations at the widest possible separations and in a wide range of unique galactic environments.

\begin{ack}[Acknowledgments]

The authors of this chapter acknowledge support by the National Aeronautics and Space Administration under grants issued through the TESS Guest Investigator Program, as well as the Section Editor for inviting them to contribute to this project.

\end{ack}

\section{Further information}

For further information on the observed population of planets orbiting giant stars, please visit http://skgrunblatt.github.io.

\seealso{Giant branch systems: dynamical and radiative evolution}

\bibliographystyle{Harvard}
\bibliography{reference}

\end{document}